\documentclass[a4paper,11pt]{article}
\pdfoutput=1

\usepackage{jheppub}

\usepackage[T1]{fontenc}
\usepackage{caption}
\usepackage{subcaption}
\usepackage{dsfont}
\usepackage{extarrows}
\usepackage[aligntableaux=center,boxsize=0.8em]{ytableau}

\allowdisplaybreaks[1]

\def\({\left(}
\def\){\right)}
\def\[{\left[}
\def\]{\right]}
\def\<{\left\langle}
\def\>{\right\rangle}

\def\be{\begin{eqnarray}}
\def\ee{\end{eqnarray}}
\def\nn{\nonumber\\}

\def\d{\delta}
\def\D{\Delta}

\def\l{\lambda}
\def\L{\Lambda}

\def\s{\sigma}

\def\nn{\nonumber\\}

\title{
Analytic trajectory bootstrap for matrix models
}

\author{Wenliang Li}
\affiliation{School of Physics, Sun Yat-Sen University, Guangzhou 510275, China}
\emailAdd{liwliang3@mail.sysu.edu.cn}

\abstract{
We revisit the large $N$ two-matrix model with $\text{tr}[A,B]^2$ interaction and quartic potentials 
by the analytic trajectory bootstrap,  
where $A$ and $B$ represent the two matrices. 
In the large $N$ limit, we can focus on the single trace moments 
associated with the words composed of 
the letters $A$ and $B$. 
Analytic continuations in the lengths of the words and subwords lead to 
analytic trajectories of single trace moments   
and intriguing intersections of different trajectories. 
Inspired by the one-cut solutions of one-matrix models, 
we propose a simple ansatz for the singularity structure of 
the two-matrix generating functions 
and the corresponding single trace moments. 
Together with the self-consistent constraints from the loop equations, 
we determine the free parameters in the ansatz  
and obtain highly accurate solutions for the two-matrix model at a low computational cost.  
For a given length cutoff $L_\text{max}$,  
our results are within and more accurate than the positivity bounds 
from the relaxation method, such as about 6-digit accuracy for $L_\text{max}=18$. 
The convergence pattern suggests that we achieve about $8$-digit accuracy for $L_\text{max}=22$.  
As the singularity structure is closely related to the eigenvalue distributions, 
we further present the results for various types of eigenvalue densities. 
In the end, we study the symmetry breaking solutions 
using more complicated ansatzes. 
}

\begin{document} 
\maketitle
\flushbottom

\section{Introduction}
Regge trajectories first appeared in a phenomenological description of hadron physics,  
but later led to string theory as a promising candidate for the quantum theory of gravity. 
Intriguingly, Regge trajectories connect particles of the same properties except for 
the angular momentum and mass. 
Can we also connect the physical observables associated with the same angular momentum by analytic continuation? 
As there is no spatial coordinate in matrix models, 
it is natural to consider this kind of analytic continuation.   
In this work, 
we study the matrix models based on analytic continuations in the numbers of matrices 
along the lines of \cite{Li:2022prn, Bender:2022eze, Bender:2023ttu, Li:2023nip, Li:2023ewe, Li:2024rod}.  

Recenly, Lin proposed a bootstrap approach for the large $N$ matrix models 
using the loop equations 
\footnote{The loop equations are special cases of the Dyson-Schwinger equations \cite{Dyson:1949ha,Schwinger:1951ex,Schwinger:1951hq} in the context of gauge theories and matrix models \cite{Makeenko:1979pb, Makeenko:1980vm, Migdal:1983qrz}.  }
and positivity requirements \cite{Lin:2020mme}.  
This method is similar in spirit to the positivity-based bootstrap methods 
for conformal field theory \cite{Rattazzi:2008pe} and lattice gauge theory \cite{Anderson:2016rcw} 
(See also \cite{Jevicki:1982jj,Jevicki:1983wu,Koch:2021yeb,Mathaba:2023non} and \cite{Hessam:2021byc,Cho:2022lcj}).  
This positive bootstrap approach can be extended to the large $N$ matrix quantum mechanics, 
in which the role of the Dyson-Schwinger equations in the Lagrangian formalism is played by 
some equations of motion in the Hamiltonian formalism \cite{Han:2020bkb}. 
\footnote{See also the bootstrap studies of quantum mechanics 
\cite{Berenstein:2021dyf,Bhattacharya:2021btd,Aikawa:2021eai,Berenstein:2021loy,Tchoumakov:2021mnh,Aikawa:2021qbl,Du:2021hfw,Lawrence:2021msm,Bai:2022yfv,Nakayama:2022ahr,Li:2022prn,Khan:2022uyz,Morita:2022zuy,Berenstein:2022ygg,Blacker:2022szo,Berenstein:2022unr,Lawrence:2022vsb,Lin:2023owt,Guo:2023gfi,Berenstein:2023ppj,Li:2023ewe,Fan:2023bld,John:2023him,Fan:2023tlh} 
and quantum many body systems \cite{Han,Nancarrow:2022wdr,Kull:2022wof,Berenstein:2024ebf}. }
As the large $N$ factorization implies that the self-consistent constraints are quadratic equations, 
it is numerically challenging to solve the resulting non-linear optimization problem. 
In \cite{Kazakov:2021lel}, Kazakov and Zheng introduced a relaxation procedure 
that treats the quadratic terms as independent variables and replaces the nonlinear equalities by linear inequalities. 
In this way, the positivity conditions can be formulated as semi-definite programming and solved more systematically. 
In \cite{Kazakov:2021lel}, they also discussed the relation between the positivity requirements in the bootstrap formulation and 
the positivity of the eigenvalue density supported only on the real axis. 
Later, this relaxation procedure was used to revisit the positive bootstrap for 
large $N$ lattice Yang-Mills theory \cite{Kazakov:2022xuh}. 
\footnote{See also the recent works \cite{Kazakov:2024ool,Li:2024wrd} on the positive bootstrap 
for finite $N$ lattice Yang-Mills theory and abelian lattice gauge theories. 
As the loop equations are linear, there is no need to use the relaxation procedure in these studies. }

In this work, we are mainly interested in the two-matrix model with $\text{tr}[A,B]^2$ interaction and quartic potential considered previously in \cite{Kazakov:2021lel}, 
which is not analytically solvable by the known methods. 
\footnote{See \cite{Kazakov:2000aq} for a review of analytically solvable matrix models. }
We want to bootstrap this two-matrix model using the analytic trajectory approach, without relying on positivity requirements.   
Therefore, our approach is sharply different from many recent positive bootstrap studies mentioned above. 
We will present highly accurate results 
based on an ansatz for the singularity structure of the generating functions. 
The challenges associated with the nonlinearity of the loop equations are partly tamed by the analytic ansatz. 
As we make use of the fully nonlinear loop equations without using the relaxation procedure, 
we are able to surpass the accuracy of the positive bootstrap results in \cite{Kazakov:2021lel}. 
Nevertheless, 
we should stress that our study is crucially guided by the high-precision positivity bounds from the relaxation bootstrap in \cite{Kazakov:2021lel}. 


The basic idea of the analytic trajectory bootstrap for matrix models is associated with the large length limit. 
This is analogous to the large spin limit in the lightcone bootstrap for conformal field theory 
\cite{Fitzpatrick:2012yx,Komargodski:2012ek}, 
where the large spin asymptotic behavior is related to the lightcone singularity.  
For matrix models, 
the large length limit of the matrix moments is also similar to the thermodynamic limit. 
In statistical mechanics, we can write the partition function of a translation invariant lattice model as
\be
Z=\text{tr}\,T^n=\sum_{\l\in S} \l^n\, \rho(\l)\xrightarrow {n\rightarrow \infty}
\l^n_{\ast}\,\rho(\l_{\ast})\,,
\label{partition-function}
\ee
where 
$T$ is the transfer matrix that relates a sequence of subsystems, 
$n$ is the number of subsystems, 
$S$ is the support of the eigenvalue distribution, 
and $\rho$ encodes the multiplicity of the discretely distributed eigenvalues
\footnote{If we take the large size limit of the subsystems, some eigenvalues of the transfer matrix can be continuously distributed. 
Here we assume that the largest eigenvalue is isolated, i.e., the theory is gapped. }.  
In the thermodynamic limit $n\rightarrow \infty$, there is an infinite number of subsystems and 
the free energy per site $F=\frac 1 V \log Z$ is dominated by 
the contribution associated with the largest eigenvalue $\l_{\ast}$. 
\footnote{We assume that 
the multiplicity or density of eigenvalues remains finite at $\l_\ast$ and all the eigenvalues are positive real.  
More generally, $\l_\ast$ denotes the eigenvalues with the largest absolute values. } 
As a landmark in theoretical physics, 
Onsager used the transfer matrix method \cite{Kramers:1941kn,Kramers:1941zz} to compute the exact free energy 
of the two-dimensional Ising model on a square lattice \cite{Onsager:1943jn}.  
It was shown that the singular critical point can emerge 
in the analytic framework of statistical mechanics by taking the thermodynamic limit. 
One can also apply the transfer matrix method to the computation of correlation functions. 
For instance, the two-point function of the two-dimensional Ising model can be written as
\be
\<S_{i_1,j_1} S_{i_2,j_2}\>&=&\frac {1} {Z}\text{tr}(\s_{i_1}^z T^{j_1-j_2}\s^z_{i_2}T^{n-j_1+j_2})
\nn&=&
\frac {1}{Z}\<\hat S_{i_1,j_1} \hat S_{i_2,j_2}T^n\>
\xrightarrow{n\rightarrow \infty}
\<0\left|\hat S_{i_1,j_1} \hat S_{i_2,j_2}\right|0\>
\,,
\label{correlation}
\ee
where $j_1>j_2$ has been assumed, $S_{i,j}$ denotes the classical spin variable, and 
$\s^z_{i}$ is the diagonal Pauli matrix associated with the $i$-th site. 
In the thermodynamic limit, 
the leading behavior of the two-point function  
can be interpreted as the vacuum expectation value of 
spin operators $\hat S_{i_1,j_1} \hat S_{i_2,j_2}$ in the Heisenberg picture, 
where the ground state is the eigenvector associated with the largest eigenvalue of the transfer matrix. 

For matrix models, it is interesting to consider the large length limit 
due to a similar mechanism of maximum eigenvalue dominance. 
\footnote{In the large $n$ limit, the significant role of the largest eigenvalues in the matrix models is similar to 
that of the eigen-energies in quantum mechanics \cite{Li:2023ewe,Li:2024rod}. 
} 
Although the basic idea is essentially the same, 
some minor modifications are needed because the density can vanish at the largest eigenvalues, 
which is different from the case of the transfer matrix. 
For two-matrix matrix models, the Green's functions of identical and mixed matrices are analogous to 
the partition function \eqref{partition-function} and correlation function \eqref{correlation}. 
In the large length limit, the leading behavior of a long Green's function is closely related 
the eigenvalues with the largest absolute values.  
The leading singularities of the generating functions should exhibit a simple structure. 
As the analogue of finite size corrections, 
the subleading terms in the large length limit should also take a relatively simple form. 
We expect that the minimally singular solutions of the two-matrix generating functions 
share some similarities with the one-cut solutions of the resolvents in the one-matrix models. 

The large length expansion of the Green's functions gives rise to asymptotic series 
that do not converge as the truncation parameter increases. 
To resolve this issue, 
we propose an analytic ansatz for the singularity structure, 
which in some sense resum the large length expansion.  
They are based on some conjectured form of the singularity structure of the generating functions. 
It turns out that they lead to rapidly convergent results even at $n=0$, 
where $n$ denotes the length of a full word or a subword. 
In this way, we are able to derive highly accurate solutions of the two-matrix model.  
Furthermore, we use a local minimization method to significantly reduce the computational cost.  

In section \ref{sec:1mm}, we revisit one-cut solutions of the quartic one-matrix model from the analytic trajectory bootstrap perspective. 
In section \ref{sec:2mm}, we use the analytic trajectory bootstrap to derive highly accurate solutions of the two-matrix model,  
which can be viewed as a natural generalization of the one-cut solutions of one-matrix models. 
In section \ref{sec:symmetry-breaking}, we extend this approach to the symmetry breaking solutions 
by introducing more complicated ansatzes. 
In section \ref{sec:discussion}, we summarize our results and discuss some directions for the future investigations. 
In Appendix \ref{sec:ansatz-integral}, we deduce the analytic ansatz 
for one-length Green's functions from the matrix integral perspective. 

\section{One-matrix model}
\label{sec:1mm}
\subsection{Resolvent and one-cut solutions}
In the large $N$ one-matrix model
\footnote{See \cite{Eynard:2015aea} for a review of various approaches to the matrix models. }
\be
Z=\lim_{N\rightarrow \infty}\int dM\, e^{-N\text{tr} V(M)}\,,
\ee 
the basic observables are the single trace moments of the Hermitian matrix $M$: 
\be
G_n=\<\text{Tr}M^{n}\>\,,
\ee
where $\text{Tr}=\frac 1 N \text{tr}$ is the normalized trace. 
For the identity matrix, we have $\text{Tr}I=1$.  
Following the standard procedure, we introduce the resolvent
\be
R(z)=\<\text{Tr}\,\frac {1}{z-M}\>
=\sum_{n=0}^\infty\,G_n\, z^{-n-1}\,.
\label{1mm-resolvent}
\ee
The matrix moments can be derived from a contour integral
\be
G_n
=
\frac{1}{2\pi i}\oint_{C} dz\,z^{n} R(z)=\int_S dz\,z^n \rho(z)\,,
\label{1mm-contour-integral}
\ee
where the contour $C$ encircles anti-clockwise all the branch points of $R(z)$. 
The support $S$ of the eigenvalue distribution and 
the eigenvalue density $\rho(z)$ are determined by  
the discontinuity of the resolvent $R(z)$. 
Usually, $n$ is assumed to be a non-negative integer, 
but it is possible to consider a complex power $n$ in the discontinuity integral. 
\footnote{In relativistic S-matrix theory and conformal field theory, analyticity in spin is established by the Froissart-Gribov formula and Caron-Hout's Lorentzian inversion formula \cite{Caron-Huot:2017vep}, respectively. 
In \eqref{1mm-contour-integral}, the contour integral is analogous to a Euclidean inversion, 
while the discontinuity integral is similar to a Lorentzian inversion. 
}
In this way, $G_n$ admits analytic continuation to complex $n$. 
See \cite{Li:2024rod} for a closely related discussion on the complex-power expectation values in the context of quantum mechanics.

For a polynomial potential $V(x)$ with $\mathbb Z_2$ symmetry, 
the symmetric one-cut solutions in Br\'ezin-Itzykson-Parisi-Zuber's classical work \cite{Brezin:1977sv} can be written as
\footnote{For symmetry breaking solutions, the building blocks for the one-cut solutions are $z^k\sqrt{(z-z_{\ast,1})(z-z_{\ast,2})}$, where $z_{\ast,1}\neq - z_{\ast,2}$. See section \ref{sec:1mm-breaking} for more details.  }
\be
R(z)=\sum_{k=0}^{k_\text{max}}\, a_k\,(z^2-z_\ast^2)^{\frac 1 2+k}
+\text{regular}\,,
\label{1mm-1cut-resolvent}
\ee
where $k_\text{max}$ depends on the degree of $V(x)$. 
The branch points are located at $z=\pm z_\ast$ due to the $\mathbb Z_2$ symmetry. 
The series expansion of a building block for the one-cut resolvent reads
\be
(z^2-z^2_\ast)^{\frac 1 2+k}&=&\sum_{j=0}^\infty \frac{(-1/2-k)_j}{j!}\, z_\ast^{2j}\,{z^{-2j+2k+1}}
\nn&=&\frac{ (-1/2-k)_{k+1}}{(k+1)!}z_\ast^{2(k+1)}\sum_{\frac n 2=-k-1}^\infty \frac{(1/2)_{n/2}}{(k+2)_{n/2}}\,z_\ast^n\,z^{-n-1}\,,
\label{1mm-buliding-block-large-z}
\ee
which can also be deduced from an integral over the support of the eigenvalue distribution (see Appendix \ref{appendix:ansatz-integral}). 
In order to match with the large $z$ expansion of the resolvent $R(z)$ in \eqref{1mm-resolvent}, 
we have changed the summation variable from $j$ to $n=2(j-k-1)$. 
The absence of odd $n$ terms is related to the $\mathbb Z_2$ symmetry constraint $\<\text{Tr}M^n\>=0$ for odd $n$. 
Therefore, we obtain an analytic ansatz for the single trace moments
\be
G_n=\frac{1+(-1)^n}{2}\sum_{k=0}^{k_\text{max}}\,b_k\,\frac{(1/2)_{n/2}}{(k+2)_{n/2}}\,z_\ast^n\,,
\label{1mm-Gn-ansatz}
\ee
where $b_k=a_k \frac{ (-1/2-k)_{k+1}}{(k+1)!}z_\ast^{2(k+1)}$. 
The $k$ summation in \eqref{1mm-Gn-ansatz} can be interpreted as a convergent expansion in large length $n$. 
The coefficients $b_k$ will be determined by the large $N$ Dyson-Schwinger equations
\be
\<\text{Tr}M^nV'(M)\>=\sum_{p=0}^{n-1}\<\text{Tr}M^p\>\<\text{Tr}M^{n-p-1}\>\,, 
\label{loop-eq-1mm}
\ee
which are called the large $N$ loop equations in the context of matrix models. 
We have omitted the subleading terms in the large $N$ limit. 
The quadratic terms on the right hand side of \eqref{loop-eq-1mm} 
come from the large $N$ factorization of double trace moments. 
The summation over the different ways of cutting the word $M^n$ can be encoded in the product of two resolvents, i.e., $R(z)^2$. 
As a result, $R(z)$ satisfies a quadratic equation and 
the branch points of $R(z)$ are of the square root type. 

\subsection{One-length trajectory bootstrap}
Let us consider the concrete example of the quartic potential
\be
V(M)=\frac 1 2 M^2+\frac g 4 M^4\,.
\ee
The loop equations read
\be
G_{n+1}+gG_{n+3}=\sum_{p=0}^{n-1}G_pG_{n-p-1}\,.
\ee
There are two free parameters. 
If we set $G_0=1$, 
the other Green's functions can be expressed in terms of $G_2$:
\be
G_4=\frac{1-G_2}{g}\,,\quad
G_6=\frac{(2g+1)G_2-1}{g^2}\,,\quad
G_8=\frac{g^2 G_2^2-(4g+1)G_2+2g+1}{g^3}\,,
\label{1mm-loop-eq-sol-1}
\ee
\be
G_{10}=\frac{-3g^2G_2^2+(6g^2+6g+1)G_2-4g-1}{g^4}\,.
\label{1mm-loop-eq-sol-2}
\ee
For symmetric one-cut solutions, 
it turns out that $k_\text{max}=1$ is sufficient for the analytic ansatz \eqref{1mm-Gn-ansatz}:
\be
G_n=\frac{1+(-1)^n}{2}\left(b_0\,\frac{(1/2)_{n/2}}{(2)_{n/2}}+b_1\,\frac{(1/2)_{n/2}}{(3)_{n/2}}\right)\,z_\ast^n\,,
\ee
which can be expressed as an integral \eqref{1mm-contour-integral} 
over the support of the eigenvalue distribution. 
The density of eigenvalues is given by
\be
\rho(z)=\frac {1}{\pi z_\ast}
\left(
2b_0\left(1-\frac{z^2}{z_\ast^2}\right)^{1/2}
+\frac {8} {3}b_1\left(1-\frac{z^2}{z_\ast^2}\right)^{3/2}
\right)\,.
\ee
Using the three relations for $(G_4, G_6, G_8)$ in \eqref{1mm-loop-eq-sol-1}, 
we can fix the coefficients and $G_2$:
\be
b_0=-1+\frac{12}{z_\ast^2}-\frac{4(9g+1)}{4+3g z_\ast^2}\,,\quad
b_1=1-b_0\,,\quad
G_2=\frac{16+gz_\ast^4}{4(4+3gz_\ast^2)}\,,
\ee
which are determined by the locations of the branch points $\pm z_\ast$. 
Although there remain infinitely many relations for ${n>8}$, 
they are not independent and imply a polynomial equation for $z_\ast^2$:
\be
3g\left(z_\ast^2\right)^2+4z_\ast^2-16=0\,. 
\ee
The two solutions of this quadratic equation read
\be
z_{\ast,\pm}^2=\frac{2}{3g}\left(-1\pm\sqrt{12g+1}\right)\,,
\ee
which give precisely the one-cut solutions for $G_2$
\be
G_2^{\text{(one-cut)},\pm}=\frac{\pm(12g+1)^{3/2}-18g-1}{54g^2}\,.
\ee
For $g>0$, the support of the eigenvalue distribution is on the real axis for the positive solution and on the imaginary axis for the negative solution. 
For $0>g>-\frac 1 {12}$, both solutions are supported on the real axis. 
As $g$ decreases, the two solutions meet at the critical point $g=-\frac 1 {12}$ and 
become a complex conjugate pair for $g<-\frac 1 {12}$. 

\section{Two-matrix model}
\label{sec:2mm}
The multi-matrix models are usually not solvable by the known analytic methods. 
As a concrete example, 
we will consider the large $N$ two-matrix model
\be
Z=\lim_{N\rightarrow \infty}\int dA dB\,
e^{-N\text{tr}V(A,B)}\,.
\ee
where the two-matrix potential reads
\be
V(A,B)=
\frac 1 2 A^2+\frac g 4 A^4+\frac 1 2 B^2+\frac g 4 B^4
-\frac {h}{2}[A,B]^2\,.
\label{2mm-definition}
\ee  
Previously, this two-matrix model has been studied using the relaxation bootstrap \cite{Kazakov:2021lel} and the Monte Carlo method \cite{Jha:2021exo}.
To make high precision comparison, we will focus on 
the $\mathbb Z_2$ symmetric solutions for $g=h=1$, 
as 6-digits results have been obtained in \cite{Kazakov:2021lel}. 

\subsection{Generating functions and one-cut solutions}
In the large $N$ two-matrix model \eqref{2mm-definition}, the basic observables are the single trace moments of identical matrices
\be
G^{(2)}_{n,0}=\<\text{Tr} \,A^{n} \>\,,\quad
G^{(2)}_{0,n}=\<\text{Tr} \,B^{n} \>\,,
\label{G2-one-length}
\ee
and the single trace moments of mixed matrices
\be
G^{(2k)}_{n_1,\,n_2,\,\dots,\,n_{2k-1},\,n_{2k}}
=\<\text{Tr}\, A^{n_1}B^{n_2}\dots A^{n_{2k-1}}B^{n_{2k}} \>\,,
\label{G2k}
\ee
where the superscript $(2k)$ indicates the number of length variables. 
For the mixed moments, we can take a contraction limit. 
For $k>1$, we have
\be
\lim_{n_j\rightarrow 0}G^{(2k)}_{\dots, n_{j-1}\,, n_{j}\,,n_{j+1}, \,\dots}
=
G^{(2k-2)}_{\,\dots, n_{j-1}+n_{j+1}, \, \dots}\,.
\label{contraction}
\ee
For $k=1$, the mixed matrix moments reduce to identical matrix moments in \eqref{G2-one-length}
\be
\lim_{n_2\rightarrow 0}G^{(2)}_{n_1,n_2}=\<\text{Tr} A^{n_1} \>\,,\quad
\lim_{n_1\rightarrow 0}G^{(2)}_{n_1,n_2}=\<\text{Tr} B^{n_2} \>\,.
\ee
As a natural generalization of the resolvents for identical matrix moments
\be
R_A(z)=\<\text{Tr}\left(\frac {1}{z-A}\right)\>\,,\quad
R_B(z)=\<\text{Tr}\left(\frac {1}{z-B}\right)\>\,,
\ee
we introduce the generating functions for the mixed moments (see e.g. \cite{Staudacher:1993xy})
\be
R^{(2k)}(z_1, \, \dots, \,  z_{2k})
&=&
\<\text{Tr}\left(\frac {1}{z_1-A}\frac {1}{z_2-B}\dots\frac {1}{z_{2k-1}-A}\frac {1}{z_{2k}-B}\right)\>
\nn&=&
\sum_{n_1=0}^\infty \dots \sum_{n_{2k}=0}^\infty z_1^{-n_1-1}\,\dots\, z_{2k}^{-n_{2k}-1}\,
G^{(2k)}_{n_1,\,\dots,\,n_{2k}}\,.
\ee
By construction, the generating functions obey some general constraints:
\be
R^{(2k)}(z_1, \, z_2, \,z_3, \, z_4, \, \dots, \, z_{2k-1}, \, z_{2k})
=
R^{(2k)}(z_3, \, z_4, \, \dots, \, z_{2k-1}, \, z_{2k},\,z_1, \, z_2)\,,
\label{R-12-34}
\ee
\be
R^{(2k)}(z_1, \, z_2, \,z_3, \dots, \, z_{2k-1}, \, z_{2k})
=
R^{(2k)}(z_1, \, z_{2k}, \,z_{2k-1}, \dots, \, z_{3}, \, z_{2})\,.
\label{R-12-12k}
\ee
If the solution is symmetric under the $\mathbb Z_2$ transformation 
$A\leftrightarrow B$, 
we further have
\be
R^{(2k)}(z_1, \, z_2, \,z_3 \, \dots, \, z_{2k})
=
R^{(2k)}(z_2, \,z_3 \, \dots, \, z_{2k},\,z_1)\,.
\label{R-12-23}
\ee
We can recover the mixed moments by evaluating the contour integral
\be
G^{(2k)}_{n_1,\,\dots,\,n_{2k}}
=
\left[\prod_{j=1}^{2k}\left(\frac{1}{2\pi i}\oint_{C} dz\,z^{n}\right)_j \right]R^{(2k)}(z_1, \, \dots, \, z_{2k})\,,
\label{2mm-contour-integral}
\ee
where the subscripts of $(z_j, C_j)$ are indicated by $(\dots)_j$.  
Since the singularity structure of $R^{(2k)}$ is related to the support of the matrix eigenvalue distribution, 
we assume that there are only branch point singularities, which are connected by branch cuts. 
The locations of the branch points in the $z_j$ complex plane are determined by the corresponding large $N$ matrix. 
The contour $C_j$ encircles all the branch points in the $z_j$ complex plane. 

For $A\leftrightarrow B$ symmetric solutions, the locations of the branch points are expected to be 
the same in all the $\{z_j\}$ complex planes. 
However, the discontinuity of $R^{(2k)}$ on the support is not necessarily a factorized function 
in $(z_1, \, \dots, \, z_{2k})$ due to the relative ``angle'' between $A$ and $B$. 
As in the one-matrix model, we focus on the $\mathbb Z_2\times\mathbb Z_2\times\mathbb Z_2$ symmetric one-cut solutions 
\be
R^{(2k)}(z_1, \dots, z_{2k})=
\sum_{k_1=0}^\infty\dots \sum_{k_{2k}=0}^\infty\,a^{(2k)}_{k_1, \dots, k_{2k}}\,
\prod_{j=1}^{2k} (z_j^2-z_\ast^2)^{\frac 1 2+k_j}
+\text{regular}\,,
\label{2mm-R-k-sum}
\ee
where the coefficients $a_{k_1, \dots, z_{2k}}$ should be compatible with 
\eqref{R-12-34}, \eqref{R-12-12k}, \eqref{R-12-23}. 
To minimize the complexity of the singularity structure \cite{Li:2023ewe}, 
we assume that the branch point singularities are essentially of the square-root type, 
i.e., they involve half-integer powers. 

By construction, the generating function $R^{(2k)}(z_1, \dots, z_{2k})$ should be regular at infinity.  
In \eqref{2mm-R-k-sum}, we assume that the branch point singularities at finite $\{z_j\}$ are encoded 
in the $\{k_j\}$ summations.  
We will further assume that the $\{k_j\}$ summations are convergent, i.e., 
their finite truncations give more and more accurate approximations for the singularity structure 
as the truncation cutoff increases. 
The ``singularity'' expansion and the large length expansion are similar in spirit to 
the lightcone bootstrap and the large spin expansion in the context of conformal field theory 
\cite{Alday:2007mf, Fitzpatrick:2012yx,Komargodski:2012ek,
Alday:2016njk,Alday:2016jfr,Simmons-Duffin:2016wlq,Caron-Huot:2017vep}. 
For the matrix models, 
we can partly resum the large length expansion due to the simpler singularity structure.  
As we will see, the $k$ summation is rapidly convergent even at small length. 

There are infinitely many generating functions and their complexity grows with the number of $\{z_j\}$ variables. In practice, we consider a finite subset of generating functions. 
In this work, we study two simple scenarios: 
\begin{enumerate}
\item 
In the first scenario, 
we consider only one-length observables as in the case of one-matrix models. 
The single-trace moments take the form 
\be
\<\text{Tr} \,{A^n\mathcal O }\>\,.
\ee 
Some simple examples are $\mathcal O=I\,, B^2\,,ABAB\,,BA^2B\,,B^4$. 
To study the analytic behavior in $n$, 
we introduce the generalized resolvent
\be
R_{\mathcal O}(z)=\<\text{Tr}\,\frac {\mathcal O}{z-A}\>
=\sum_{n=0}^\infty \<\text{Tr} \,\mathcal O A^{n}\> z^{-n-1}\,,
\label{generalized-resolvent}
\ee
which are related to the generating function $R^{(2k)}(z_1, \dots, z_{2k})$ by certain contour integrals. 
The standard resolvent corresponds to the case of the identity operator $I$. 
As in \eqref{2mm-R-k-sum}, 
we assume that the one-cut solution of a generalized resolvent 
involves an infinite sum of $(z^2-z_\ast^2)^{\frac 1 2+k}$ with $k=0,1,2,\dots$. 
After truncating the $k$ summation, a generalized resolvent is approximated by
\be
R_{\mathcal O}(z)\approx\sum_{k=0}^{k_\text{max}}\,a_{k, \mathcal O}\,(z^2-z_\ast^2)^{\frac 1 2+k}
+\text{regular}\,.
\ee
Therefore, the analytic ansatz for the one-length Green's functions is given by
\be
\<\text{Tr}\,\mathcal O A^{n}\>\approx\sum_{k=0}^{k_\text{max}}\,b_{k,\mathcal O}\,\frac{(1/2)_{n/2}}{(k+2)_{n/2}}\,z_\ast^n\,,
\label{1-length-ansatz}
\ee
According to the $A\rightarrow -A$ symmetry, there exists an implicit factor, depending on the number of $A$ in $\mathcal O$. 
If $\mathcal O$ contains an even number of $A$, the factor is $\frac{1+(-1)^n}{2}$. 

\item 
In the second scenario, we focus on the simplest multi-length observables 
\be
G^{(2)}_{n_1,n_2}=\<\text{Tr}\,A^{n_1}B^{n_2}\>\,,
\ee 
which contain two length variables.  
They reduce to the identical matrix moments in a contraction limit $n_j\rightarrow 0$. 
In the approximate one-cut solutions,  
the $k_j$ summations in the two-length generating function are truncated to finite sums:
\be
R^{(2)}(z_1, z_2)\approx
\sum_{k_1=0}^{k_\text{max}}\,\sum_{k_2=0}^{k_\text{max}-k_1}\,a^{(2)}_{k_1,k_2}\,
(z_1^2-z_\ast^2)^{\frac 1 2+k_1}(z_2^2-z_\ast^2)^{\frac 1 2+k_2}
+\text{regular}\,.
\ee
Accordingly, the analytic ansatz for the two-length Green's functions reads
\be
G_{n_1,n_2}\approx
\sum_{k_1=0}^{k_\text{max}}\,\sum_{k_2=0}^{k_\text{max}-k_1}\,b_{k_1,k_2}\,
\frac{(1/2)_{{n_1}/{2}}\,(1/2)_{{n_2}/{2}}}
{(k_1+2)_{{n_1}/{2}}\,(k_2+2)_{{n_2}/{2}}}\,z_\ast^{n_1+n_2}\,,\qquad
\label{2mm-analytic-ansatz-G}
\ee
where the $A\leftrightarrow B$ symmetry implies $b_{k_1,k_2}=b_{k_2,k_1}$. 
We do not write the superscript ${(2)}$ and the factors
$\frac{1+(-1)^{n_1}}{2}\frac{1+(-1)^{n_2}}{2}$ explicitly for notational simplicity. 
The additional factors are associated with the invariance under 
the transformations $A\rightarrow -A$ and $B\rightarrow -B$.
\end{enumerate}
As in the one-matrix model examples, 
the free parameters will be determined by the large $N$ loop equations
\be
\<\text{Tr}\,\mathcal O[A+gA^3-h(2BAB-ABB-BBA)]\>
=\sum_{p=1}^n \<\text{Tr}\mathcal O^{(A,l)}_{p-1}\>\< \text{Tr}\mathcal O^{(A,r)}_{n-p}\>\,,
\label{loop-eq-2mm}
\ee
where $n$ denotes the length of the subword $\mathcal O$. 
The loop equations associated with the $B$ derivative is redundant due to the $A\leftrightarrow B$ symmetry. 
As a two-matrix generalization of \eqref{loop-eq-1mm}, 
the right hand side of \eqref{loop-eq-2mm} is summing over all possible ways of splitting $\mathcal O$ 
according to the position of $A$, i.e., 
the $p$-th summand vanishes if the $p$-th letter of $\mathcal O$ is $B$. 
We use the superscript $l$ or $r$ to indicate the left or right subword. 
See the Appendix E of \cite{Kazakov:2021lel} for some concrete examples of the loop equations for short words.

The loop equations in \eqref{loop-eq-2mm} imply that the Green's functions are related to each other. 
It turns out that only a small number of them are free parameters. 
All the Green's functions can be expressed in terms of the two-length Green's functions $G_{n_1,n_2}=\<\text{Tr}\,A^{n_1}B^{n_2}\>$, where the one-length Green's functions $G_{n}=\<\text{Tr}\,A^{n}\>=\<\text{Tr}\,B^{n}\>$ are the special cases with $n_1=0$ or $n_2=0$. 
We choose the expectation values of the one-length words and the simplest type of two-length words 
as the independent parameters:
\be
G_{4j+2}, \quad G_{4j+4},\quad G_{4j+2,2}\,,
\label{2mm-independent-G}
\ee
where $j=0,1,2,\dots$. 
\footnote{If we set $g=h=1$ before solving the loop equations, 
it may appear that there are additional independent parameters for long words. 
Since the solutions of the loop equations should be analytic in $g$ and $h$ around $g=h=1$, 
we further impose that the loop equations are solved for $h=1+\d h$ to first order in $\d h$. 
In this way, we eliminate the additional independent parameters by the analyticity requirement.  
All Green's functions are determined by the independent ones in \eqref{2mm-independent-G}. 
For instance, we find one more free parameter at $L_\text{max}=12$ and fix it by solving the loop equations to first order in $\d h=h-1$.   }
For notational simplicity, we will not write the superscript of the Green's functions explicitly.
The normalized trace indicates $G_0=1$. 
Some explicit solutions associated with the short words are
\be
G_{1,1,1,1}=\frac 1 2 \left(-1+G_2+G_4+2G_{2,2}\right)\,,
\quad
G_{2,4}=\frac 1 6 \left(1+G_2-6G_{2,2}+G_6\right)\,,
\ee
\be
G_{1,1,1,3}=\frac 1 6\left(1-5G_2+3 G_4-6G_{2,2}+4 G_6\right)\,,
\quad
G_{1,2,1,2}=-G_2+G_4-G_{2,2}+G_6\,,\qquad
\ee
\be
G_{1,1,1,5}=\frac 1 2 (-(G_2)^2-2G_4+G_6+G_8+2G_{2,6})\,,
\ee
\be
G_{1,2,1,4}=\frac 1 3(1-8G_2-14(G_2)^2-G_4-14G_{2,2}+13G_6+6G_8+12G_{2,6})\,,
\ee
\be
G_{4,4}=\frac 1 6(-3+15G_2+22(G_2)^2-4G_4+34G_{2,2}-21G_6-6G_8-18G_{2,6})\,.
\label{G44}
\ee
We notice that a Green's function of length-$n$ can be expressed in terms of the independent ones in \eqref{2mm-independent-G} of at most length-$n$. 
For $n\geq 8$, the solutions contain quadratic terms in the independent parameters. 
We also find higher order terms for larger length-$n$, such as cubic terms for $n=14$ and quartic terms for $n=20$. 

In the first scenario, 
some intricate matching conditions for the one-length ansatz arise at the trajectory intersections 
due to the contraction limits and the symmetries of the Green's functions. 
\footnote{
In 2D conformal field theory, the Regge trajectories are associated with the analytic continuation in the $SL(2,\mathbb R)$ conformal spin. 
Since the spectrum is symmetric in the conformal spin $\D+\ell$ and twist $\D-\ell$, 
one can also consider the analytic continuation in twist. 
The intersections of the two types of analytic trajectories give rise to the rigid Virasoro structure \cite{Li:2021uki}. 
Here we show that the intersections of different analytic trajectories also appear in the context of matrix models, 
which can be viewed as a generalization of the Virasoro intersections. 
The existence and intersections of analytic trajectories provide a general mechanism for the emergence of rigid structures, 
which can be useful in diverse bootstrap contexts. }
Some of them can be automatically taken into account by the two-length ansatz in the second scenario. 
As we will see, the second scenario leads to a more systematic procedure and 
highly accurate solutions for the two-matrix model \eqref{2mm-definition}.  

\subsection{One-length trajectory bootstrap}
\begin{figure}[h]
	\centering
		\includegraphics[width=0.9\linewidth]{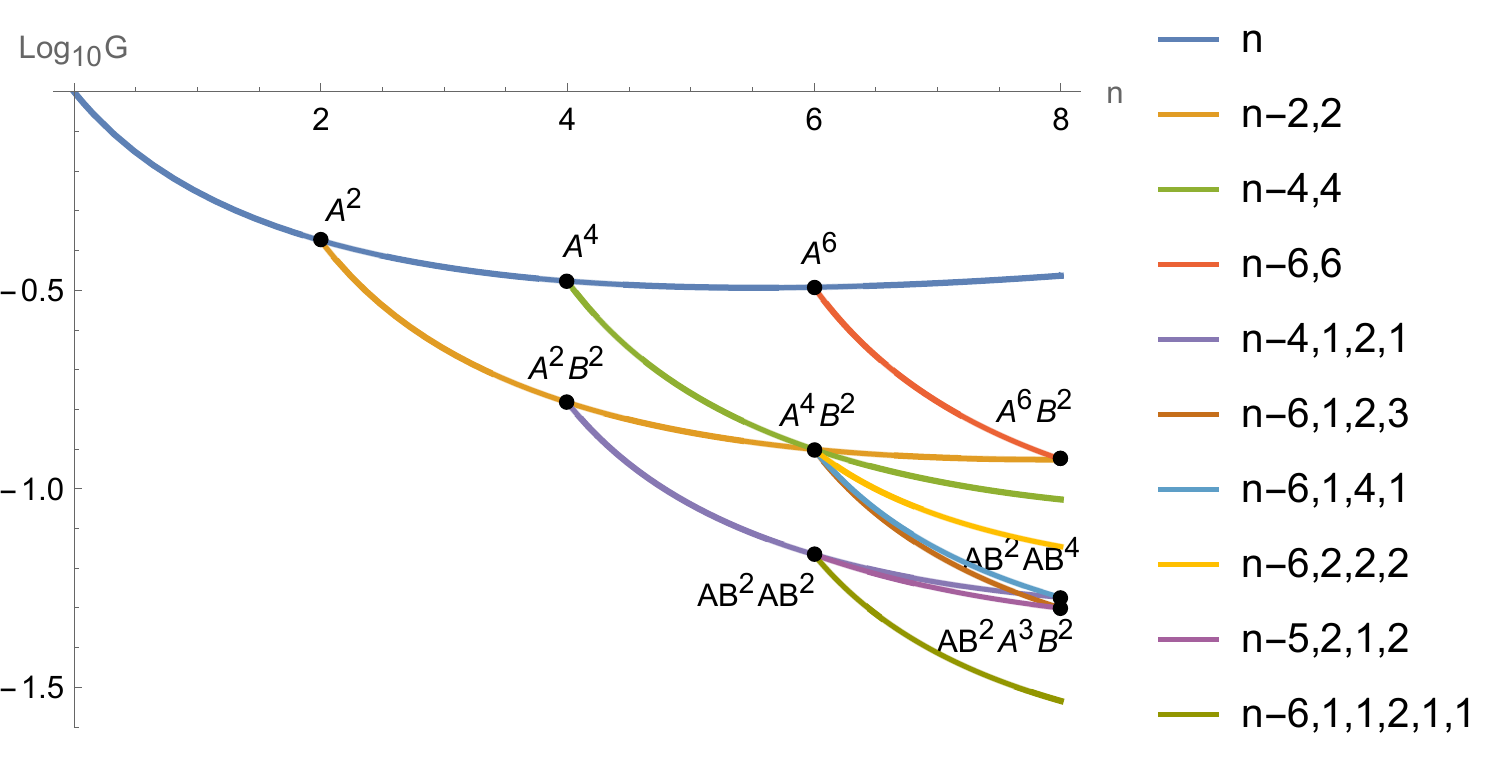}
	\caption{The analytic trajectories associated with the Green's functions 
	$\<\text{Tr}\,A^{n-L_\mathcal O}\mathcal O \>$ of the two-matrix model \eqref{2mm-definition}. 
	Here $n$ is the length of the full word $ A^{n-L_\mathcal O}\mathcal O$, and $L_\mathcal O$ is the length of the subword $\mathcal O$. 
	The trajectories are labelled by the powers of the letters as in \eqref{G2k}. 
	According to the contraction limit \eqref{contraction} and symmetries of the Green's functions, 
	these analytic trajectories exhibit intriguing intersection phenomena, 
	which lead to nontrivial matching conditions for the one-length analytic ansatz \eqref{1-length-ansatz}. 
	We write down some explicit words associated with the intersection points. 
	For clarity, we have stripped off the common factor $(1+(-1)^n)/2$ associated with $\mathbb Z_2$ symmetry.  }
	\label{fig:one-length-trajectories}
\end{figure}
The one-length trajectories exhibit intriguing intersection phenomena. 
The intersections can be associated with the contraction limit, such as
\be
\<\text{Tr}\,A^2\>=\<\text{Tr}\,A^2B^0\>\,,\quad
\<\text{Tr}\,BA^2B\>=\<\text{Tr}\,BA^2BA^0\>\,,
\ee
They can also be related to the symmetries of the Green's functions, such as 
\be
\<\text{Tr}\,A^2B^2\>=\<\text{Tr}\,B^2A^2\>=\<\text{Tr}\,BA^2B\>\,.
\ee
Assuming the invariance under the transformation $A\leftrightarrow B$, 
we can always use $\<\text{Tr}\,A^n\mathcal O \>$ to represent a one-length trajectory. 
Some examples for the trajectory intersections are:
\begin{itemize}
\item  Intersection at $n=2$
\be
\<\text{Tr}\,A^n\>=\<\text{Tr}\,A^{n-2}B^2\>\,.
\ee
\item Intersections at $n=4$
\be
\<\text{Tr}\,A^n\>=\<\text{Tr}\,A^{n-4}B^4\>\,,\quad
\<\text{Tr}\,A^{n-2}B^2\>=\<\text{Tr}\,A^{n-4}BA^2B\>\,.
\qquad
\ee
\item Intersections at $n=6$
\be
\<\text{Tr}\,A^n\>=\<\text{Tr}\,A^{n-6}B^6\>\,,\quad
\ee
\be
\<\text{Tr}\,A^{n-2}B^2\>
&=&\<\text{Tr}\,A^{n-4}B^4\>
=\<\text{Tr}\,A^{n-6}BA^4B\>
\nn&=&
\<\text{Tr}\,A^{n-6}B^2A^2B^2\>
=\<\text{Tr}\,A^{n-6}BA^2B^3\>\,,
\ee
\be
\<\text{Tr}\,A^{n-4}BA^2B\>
=
\<\text{Tr}\,A^{n-5}B^2AB^2\>
=
\<\text{Tr}\,A^{n-6}BAB^2AB\>\,.
\ee
\end{itemize}
Here we write the explicit Green's functions in terms of the letters $A$ and $B$. 

In figure \ref{fig:one-length-trajectories}, we present the one-length trajectories as analytic functions 
of the length of the complete word, which is based on the solutions for $(L_\text{max}, k_\text{max})=(12,2)$. 
We can see more explicitly the intersection phenomena described above. 
We restrict to a subset of trajectories that are connected to 
$G_{0,0}=\<\text{Tr}\,I\>=1$. 
The maximal word length is $n_\text{max}=8$. 
We use the more concise notation introduced in \eqref{G2k} to denote 
various one-length trajectories of the Green's functions.

\begin{table}[h]
\center
\begin{tabular}{c c c c c c }
\hline\hline
$L_\text{max}$ & $k_\text{max}$ & $z_\ast^2$ & $\<\text{Tr} A^2\>$  &$\<\text{Tr} A^4\>$ &$\<\text{Tr} A^2B^2\>$  \\
\hline 
$6$    & 0	&$1.5769834$ & $0.39424586$ & $0.31085960$ & $0.15542980$\\
$8$  	 & 1	&$1.4940026$ & $0.41986057$ & $0.33095185$ & $0.16422691$\\
$10$  & 1	&$1.4942817$ & $0.42341198$ & $0.33496772$ & $0.16508449$\\
$10$  & 2	&$1.4939148$ & $0.42308874$ & $0.33464996$ & $0.16506880$\\
$12$  & 1	&$1.4930493$ & $0.41616840$ & $0.32669520$ & $0.15533998$\\
$12$  & 2	&$1.4960073$ & $0.42178579$ & $0.33333764$ & $0.16495112$\\
\hline\hline
\end{tabular}
\caption{The one-length trajectory bootstrap results of the two-matrix model \eqref{2mm-definition} with $g=h=1$, where we present 8 significant digits for the largest eigenvalue square and the independent Green's functions of length-2 and length-4. 
More accurate results from the two-length trajectory bootstrap are presented in table \ref{table-2mm}. 
 }
\label{table-2mm-1-length}
\end{table}

In table \ref{table-2mm-1-length}, we summarize the results of the one-length trajectory bootstrap for $L_\text{max}\leq 12$. 
For a given length cutoff $L_\text{max}$, we choose a truncation order $k_\text{max}$ for the analytic ansatz in \eqref{1-length-ansatz}. 
Then we solve the matching conditions  associated with the trajectory intersections and 
minimize the $\eta$ function
\be
\eta&=&
(G_0-1)^2+
\left(G_{1,1,1,1}-\frac 1 2 \left(-1+G_2+G_4+2G_{2,2}\right)\right)^2
\nn&&+
\left(G_{2,4}-\frac 1 6 \left(1+G_2-6G_{2,2}+G_6\right)\right)^2
+\dots\,,
\ee
which is constructed from the loop equations solutions with length not greater than $L_\text{max}$ and the Green's functions are replaced by \eqref{1-length-ansatz}. 
If $k_\text{max}$ is too small, we are not able to solve the matching conditions associated with the trajectory intersections. 
If $k_\text{max}$ is too large, 
the number of free parameters is greater than the number of constraints 
and the system is underdetermined, 
then we cannot solve for $z_\ast^2$ and the coefficients in the ansatz for the one-length trajectories. 
Therefore, we find only one or two choices for $k_\text{max}$ for a given $L_\text{max}\leq 12$. 
For the same $L_\text{max}$, the accuracy increases with the truncation order $k_\text{max}$ of the analytic ansatz. 

As $L_\text{max}$ increases, there are more matching conditions from the intricate intersections of different trajectories. 
Furthermore, the number of coefficients in the analytic ansatz also increases rapidly. 
As a result, the computational complexity grows rapidly. 
Below, we will consider the two-length trajectory bootstrap, 
whose results can be improved in a more systematic way and at a lower computational cost. 

\subsection{Two-length trajectory bootstrap}
Above, the matching conditions associated with the intersections of the one-length trajectories 
are imposed one by one.  
In fact, these constraints are encoded in the multi-length generating functions $R^{(2k)}(z_1, \, \dots, \,  z_{2k})$.   
As different one-length trajectories are unified by the multi-length trajectories,  
analyticity guarantees that the results are independent of the path of analytic continuations. 
Below, we consider the simplest multi-length generating functions, 
i.e. $R^{(2)}(z_1,z_2)$ for the two-length Green's functions. 
At the price of one more variable, the matching conditions $\<\text{Tr}\,A^{n_1}\>=\<\text{Tr}\,A^{n_1}B^{n_2}\>|_{n_2\rightarrow 0}$ can be satisfied automatically. 
According to the pattern of the independent parameters in \eqref{2mm-independent-G}, 
it is natural to set the length cutoff as $L_\text{max}=4p+6$, where $p$ is a non-negative integer. 
For a specific $k_\text{max}$ in the analytic ansatz \eqref{Gnn-kmax-0},  
the number of free parameters in \eqref{2mm-analytic-ansatz-G} 
is the same as that of constraints for the two-length Green's functions with the length cutoff $L_\text{max}= 2k_\text{max}+6$, 
so we set $k_\text{max}=2p$.
\footnote{In fact, we also examine the cases of $k_\text{max}=1,3,5,7$.  
The results also converge to our highly accurate solution at $k_\text{max}=8$.   
However, their convergence rate is lower than the cases of even $k_\text{max}$ 
and obscures the rapid convergence pattern of even $k_\text{max}$. } 
Below, we consider the cases of $p=0,1,2,3,4$, 
i.e., $L_\text{max}=6,10,14,18,22$. 

\subsubsection{$L_\text{max}=6$}
For $L_\text{max}=6$, the constraints for the two-length Green's functions are
\be
G_{0,0}=1\,,\quad
G_{2,4}=\frac 1 6 \left(1+G_2-6G_{2,2}+G_6\right)\,,
\label{2L-constraints-0}
\ee
where the first one is the normalization condition and the second one is derived from the loop equations \eqref{loop-eq-2mm}. 
This corresponds to the lowest truncation order $k_\text{max}=0$, 
so the analytic ansatz for two-length Green's functions takes a factorized form
\be
\<\text{Tr}\,A^{n_1}B^{n_2}\>
=b_{0,0}\frac{(1/2)_{{n_1}/{2}}\,(1/2)_{{n_2}/{2}}}
{(2)_{{n_1}/{2}}\,(2)_{{n_2}/{2}}}\,z_\ast^{n_1}z_\ast^{n_2}\,,
\label{Gnn-kmax-0}
\ee
where $z_\ast^2$ and $b_{0,0}$ are free parameters. 
The constraints \eqref{2L-constraints-0} imply $b_{0,0}=1$ and a degree-3 polynomial equation for $z_\ast^2$:
\be
\quad
7\left(z^2_\ast\right)^3+24\left(z^2_\ast\right)^2-16\left(z_\ast^2\right)-64=0\,.
\ee 
There are three solutions for $z_\ast^2$. Only one solution is positive real and the other two solutions are negative real. 
In this work, we are mainly interested in the solution with eigenvalue support on the real axis, 
so we focus on the positive solution, which reads 
\be
z_\ast^2\approx1.59552\,.
\label{zs-L6}
\ee 
Accordingly, the short Green's functions from the analytic ansatz \eqref{Gnn-kmax-0} are
\be
\<\text{Tr} A^2\>=\frac{z_\ast^2}{4}\approx0.39888\,,\quad
\<\text{Tr} A^4\>=\frac{z_\ast^4}{8}\approx0.31821\,,\quad
\<\text{Tr} A^2B^2\>=\frac{z_\ast^4}{16}\approx0.15911\,,\qquad
\label{2mm-2length-Lmax6}
\ee
which are closely related to each other 
due to the simple form of the ansatz in \eqref{Gnn-kmax-0}. 
Our $L_\text{max}=6$ result for $\<\text{Tr}  A^2\>$ in \eqref{2mm-2length-Lmax6} is consistent with 
the positivity bounds for $L_\text{max}=8$ from the relaxation method in \cite{Kazakov:2021lel}: 
\be
0.393566\leq \<\text{Tr} A^2\>\leq 0.431148\,.
\label{L-8-bounds}
\ee 

\subsubsection{$L_\text{max}=10$}
To increase the truncation order  for the analytic ansatz to $k_\text{max}=2$,  
we set the length cutoff as $L_\text{max}=10$. 
The free parameters in the analytic ansatz \eqref{2mm-analytic-ansatz-G} are
\be
z_\ast^2\,,\quad
b_{0,0}\,,\quad
b_{0,1}\,,\quad
b_{0,2}\,,\quad
b_{1,1}\,.
\ee
For $L_\text{max}=10$, 
the self-consistent constraints for the Green's functions are associated with 
\be
G_{0,0}\,,\quad
G_{2,4}\,,\quad
G_{4,4}\,,\quad
G_{2,8}\,,\quad
G_{4,6}\,. 
\ee
We can see explicitly that the numbers of free parameters and constraints do match. 
The explicit constraints are \eqref{2L-constraints-0}, \eqref{G44}, and
\be
G_{2,8}&=&\frac 1 {450}\big(-52+572G_2+84G_4+876G_{2,2}
-757G_6-378G_8-1206G_{2,6}-45G_{10}
\nn&&\qquad
+G_2(849G_2+426G_4+252G_{2,2})
\big)\,,
\ee
\be
G_{4,6}&=&\frac 1 {30}\big(11-61G_2+8G_4-128G_{2,2}+86G_6+54G_8+48G_{2,6}+15G_{10}
\nn&&\qquad-G_2(107G_2+18G_4+6G_{2,2})\big)\,. 
\ee
They lead to nontrivial constraints for the independent Green's functions with $L_\text{max}=10$ in \eqref{2mm-independent-G}. 
In this case, we obtain a degree-25 polynomial equation for $z_\ast^2$, 
and the coefficients $\{b_{k_1,k_2}\}$ are given by degree-24 polynomials in $z_\ast^2$. 
\footnote{We use Mathematica's \texttt{GroebnerBasis} to derive the degree-25 polynomial equation in $z_\ast^2$,  and the expressions of $b_{k_1,k_2}$ in terms of degree-24 polynomials in $z_\ast^2$.  }
Among the 25 solutions, 
we find seven pairs of complex conjugate solutions and eleven real solutions,  
but only two of them are positive real:
\be
z_\ast^2\approx1.49915\,,\quad 
2.13966\,.
\label{zs-L10}
\ee 
According to the $L_\text{max}=6$ result, the first solution seems more reliable. 
We will focus on the solution around $z_\ast^2\approx 1.5$ below. 
Then we further determine the coefficients $b_{k_1,k_2}$ and the Green's functions, such as
\be
\<\text{Tr} A^2\>\approx0.423552\,,\quad
\<\text{Tr} A^4\>\approx0.334978\,,\quad
\<\text{Tr} A^2B^2\>\approx0.165201\,,
\ee
which should be more accurate than those in \eqref{2mm-2length-Lmax6}. 
As expected, our $L_\text{max}=10$ determination of $\<\text{Tr} A^2\>$ is also 
within the $L_\text{max}=8$ bounds in \eqref{L-8-bounds}.
Since the positivity bounds for $L_\text{max}=10,12$ are not presented in \cite{Kazakov:2021lel}, 
we are not able to make more comparisons with the positivity bounds from the relaxation bootstrap.

\subsubsection{$L_\text{max}=14, 18, 22$}
It is straightforward to apply the above procedure to higher truncation orders.  
We first need to solve the longer loop equations and express the two-length Green's functions in terms of the independent variables in \eqref{2mm-independent-G}. 
As the number of loop equations grow rapidly with $L_\text{max}$, 
this step becomes computationally more and more expensive. 
After solving the loop equations, we use the relations between the two-length Green's functions to determine the free parameters in the analytic ansatz \eqref{2mm-analytic-ansatz-G}. 
We will use a local minimization method to extract the target solution of a set of polynomial equations, 
which is applicable to non-positive solutions and 
computationally cheaper than the positivity-based optimization methods.  
Our approach will not give rigorous bounds, so it is more difficult to estimate the errors in the results.  

As the length cutoff $L_\text{max}$ increases, we need to solve a large system of high degree polynomial equations, 
so it becomes more and more challenging to obtain all the solutions. 
Using the trick in \cite{Li:2024rod}, we transform this difficult problem 
into a local minimization problem by introducing the $\eta$ function
\be
\eta&=&\left(G_{0,0}-1\right)^2+\left(G_{2,4}-\frac{1+G_2-6G_{2,2}+G_6}{6}\right)^2
\nn&&
+\Big(G_{4,4}-\frac 1 6(-3+15G_2+22G_2^2-4G_4+34G_{2,2}-21G_6-6G_8-18G_{2,6})\Big)^2
\nn&&
+\dots\,,
\label{eta-function}
\ee
which encodes the normalization condition and 
the self-consistent constraints for the two-length Green's functions from the loop equations, such as those in \eqref{2L-constraints-0} and \eqref{G44}. 
In \eqref{eta-function}, we should substitute $G_{n_1,n_2}$ with the analytic ansatz \eqref{2mm-analytic-ansatz-G}, 
so $\eta$ is a function of the largest eigenvalue squared $z_\ast^2$ 
and the coefficients $b_{k_1,k_2}$ of the building blocks in \eqref{2mm-analytic-ansatz-G}. 
The $z_\ast^2$ dependence is highly nonlinear. 
However, we find it easy to minimize the $\eta$ function at a fixed $z_\ast^2$ 
despite the nonlinear dependence on $b_{k_1,k_2}$. 

For $L_\text{max}=14$, the positive real solutions for $z_\ast^2$ of the polynomial system are $z_\ast^2\approx 1.496, 1.694, 2.360, 5.205$. 
The first solution at $L_\text{max}=14$ can be identified with the first solutions 
in \eqref{zs-L6}, \eqref{zs-L10} at lower truncation orders. 
Based on the one-length trajectory bootstrap results in table \ref{table-2mm-1-length} 
and the two-length trajectory bootstrap results for $L_\text{max}=6,10,14$, 
we conjecture that the exact solution is located around $z_\ast^2\approx 1.5$. 
Accordingly,  
we find only one local minimum around $z_\ast^2\approx 1.5$ at the higher length cutoff $L_\text{max}=18,22$ 
and obtain rapidly convergent results. 
As we are mainly interested in the solutions with $z_\ast^2\approx 1.5$, 
the starting values for the local minimization should be chosen properly. 
In practice, we first determine the starting values of $b_{k_1,k_2}$ by minimizing the $\eta|_{z^2_\ast= 1.5}$ function where $z_\ast^2$ is near the conjectured exact value. 
Then we carry out a local minimization with $1.5$ as the starting value for $z_\ast^2$.  
In spite of high nonlinearity, we always obtain a local minimum 
with an extremely small $\eta_{\text{min}}$, which approaches zero as the working precision increases. 
Then we substitute the $\eta$ minimization results into the analytic ansatz \eqref{2mm-analytic-ansatz-G}.  
This allows us to evaluate the two-length Green's functions 
and thus all single-trace moments of the target solution. 

\begin{table}[h]
\center
\begin{tabular}{c c c c c c }
\hline\hline
$L_\text{max}$ & $k_\text{max}$ & $z_\ast$ & $\<\text{Tr} A^2\>$  &$\<\text{Tr} A^4\>$ &$\<\text{Tr} A^2B^2\>$  \\
\hline 
$6$    & 0	&$\underline{1.2}63139202$ & $\underline{0}.3988801608$ & $\underline{0.3}182107654$ & $\underline{0.1}591053827$\\
$10$  & 2	&$\underline{1.22}4398880$ & $\underline{0.42}35522795$ & $\underline{0.33}49775096$ & $\underline{0.16}52013401$\\
$14$  & 4	&$\underline{1.2230}20278$ & $\underline{0.4217}471346$ & $\underline{0.3333}103568$ & $\underline{0.1649}495485$\\
$18$  & 6	&$\underline{1.22304}0962$ & $\underline{0.42178}35573$ & $ \underline{0.333341}3331$ & $\underline{0.1649530}130$\\
$22$  & 8	&$1.223041990$ & $0.4217840726$ & $0.3333417063$ & $0.1649530476$\\
\hline\hline
\end{tabular}
\caption{The two-length trajectory bootstrap results for the largest eigenvalue $z_\ast$ and some independent  Green's functions of the two-matrix model \eqref{2mm-definition} with $g=h=1$.  
The length cutoff $L_\text{max}$ restricts the number of constraints from 
the loop equation \eqref{loop-eq-2mm}. 
The results for $\<\text{Tr} A^2\>$ and $\<\text{Tr} A^4\>$ are within the positivity bounds 
at higher $L_\text{max}$ in \cite{Kazakov:2021lel}. 
We underline the stable digits for $L_\text{max}=6,10,14,18$. 
According to the convergence pattern,  
the results are of about 8-digit accuracy for $L_\text{max}=22$.
 }
\label{table-2mm}
\end{table}

In table \ref{table-2mm}-\ref{table-2mm-5}, 
we summarize the results for the largest eigenvalue $z_\ast$ and some Green's functions 
at $L_\text{max}=6,10, 14, 18, 22$. 
We can see that the results converge rapidly as $L_\text{max}$ increases, 
including the Green's functions of length-32. 
In the positivity bound approach, it requires more efforts to determine the Green's functions with larger lengths, so they were not presented in \cite{Kazakov:2021lel}. 
In our approach, it is straightforward to make predictions for the longer Greens' functions using the analytic ansatz \eqref{2mm-analytic-ansatz-G}, which covers all the independent parameters in \eqref{2mm-independent-G} and thus all the single trace moments through the loop equations. 
Furthermore, we find that the results for the longer Green's functions also converge rather rapidly, 
which is related to the dominance of the contributions around the largest eigenvalues in the matrix integral. 

For comparison, we summarize the rigorous bounds  
from the relaxation formulation of the positive bootstrap \cite{Kazakov:2021lel}:
\be
0.421780275 \leq \< \text{Tr} A^2 \>\leq 0.421785491\,,\quad
0.333339083 \leq \< \text{Tr} A^4 \>\leq 0.333343006\,,\quad
\label{relaxation-L20}
\ee
\be
0.421783612 \leq \< \text{Tr} A^2 \>\leq 0.421784687\,,\quad 
0.333341358 \leq \< \text{Tr} A^4 \>\leq 0.333342131\,,\quad
\label{relaxation-L22}
\ee
where \eqref{relaxation-L20} is for $L_\text{max}=20$ and \eqref{relaxation-L22} is for $L_\text{max}=22$. 
\footnote{The parameter $\L$ in \cite{Kazakov:2021lel} is related to the length cutoff by $L_\text{max}=2\L$.}
Our results for $\<\text{Tr} A^2\>$ and $\<\text{Tr} A^4\>$ at $L_\text{max}=14,18$ are within the positivity bounds in \cite{Kazakov:2021lel} with $L_\text{max}=16,20$. 
In fact, our $L_\text{max}=18$ results are just slightly below the lower bounds in \eqref{relaxation-L22},
so they already have similar accuracy as that of the positivity bounds at the higher length cutoff $L_\text{max}=22$ in \cite{Kazakov:2021lel}. 

\begin{table}[h]
\center
\begin{tabular}{c c c c c c c}
\hline\hline
$L_\text{max}$  & $\<\text{Tr} A^6\>$ &$\<\text{Tr} A^8\>$ &$\<\text{Tr} A^6B^2\>$ \\
\hline 
$6$    & $\underline{0.3}173199032$ & $\underline{0.3}544033193$ & $\underline{0.1}265726140$\\
$10$  & $\underline{0.32}34044907$ & $\underline{0.34}59107152$ & $\underline{0.11}91048039$\\
$14$  & $\underline{0.3216}387230$ & $\underline{0.343}7807762$ & $\underline{0.1185}379754$\\
$18$  & $\underline{0.32167}19741$ & $\underline{0.343822}3591$ & $\underline{0.118550}0944$\\
$22$  & $0.3216724315$ & $0.3438229503$ & $0.1185502550$\\
\hline\hline
\end{tabular}
\caption{The two-length trajectory bootstrap results for the independent Green's functions of length-$6, 8$. 
Using the analytic ansatz \eqref{2mm-analytic-ansatz-G}, we also make predictions for the Green's functions 
with lengths greater than the length cutoff $L_\text{max}$ for the self-consistent constraints.
 }
\label{table-2mm-2}
\end{table}

\begin{table}[h]
\center
\begin{tabular}{c c c c c c c}
\hline\hline
$L_\text{max}$  
&$\<\text{Tr} A^{10}\>$ &$\<\text{Tr} A^{12}\>$ &$\<\text{Tr} A^{10}B^2\>$ \\
\hline 
$6$   & $\underline{0}.4240933591$ & $\underline{0}.5316533429$ & $\underline{0.1}691624273$\\
$10$ & $\underline{0.39}40839495$ & $\underline{0.46}86863738$ & $\underline{0.14}18586490$\\
$14$ & $\underline{0.391}2843876$ & $\underline{0.4648}134622$ & $\underline{0.1405}385457$\\
$18$ & $\underline{0.39134}04816$ & $\underline{0.46489}19176$ & $\underline{0.14055}88137$\\
$22$ & $0.3913412639$ & $0.4648929777$ & $0.1405590861$\\
\hline\hline
\end{tabular}
\caption{The two-length trajectory bootstrap results for the independent Green's functions of length-$10, 12$. 
 }
\label{table-2mm-3}
\end{table}

\begin{table}[h]
\center
\begin{tabular}{c c c c c c c}
\hline\hline
$L_\text{max}$  
&$\<\text{Tr} A^{14}\>$ &$\<\text{Tr} A^{16}\>$ &$\<\text{Tr} A^{14}B^2\>$ \\
\hline 
$6$   & $\underline{0}.6892144055$ & $\underline{0}.9163798430$ & $\underline{0.2}749139529$\\
$10$ & $\underline{0.5}750990741$ & $\underline{0.7}226379655$ & $\underline{0.20}44859817$\\
$14$ & $\underline{0.569}5818537$ & $\underline{0.714}6463513$ & $\underline{0.2016}429636$\\
$18$ & $\underline{0.56969}35852$ & $\underline{0.71480}70340$ & $\underline{0.20167}68661$\\
$22$ & $0.5696950513$ & $0.7148090950$ & $0.2016773715$\\
\hline\hline
\end{tabular}
\caption{The two-length trajectory bootstrap results for the independent Green's functions of length-$14, 16$. 
Our results for $\<\text{Tr} A^{16}\>$ are compatible with and more accurate than the Monte Carlo result $0.7153(8)$ in \cite{Jha:2021exo}.
 }
\label{table-2mm-4}
\end{table}

\begin{table}[h]
\center
\begin{tabular}{c c c c c c c}
\hline\hline
$L_\text{max}$  
&$\<\text{Tr} A^{30}\>$ &$\<\text{Tr} A^{32}\>$ &$\<\text{Tr} A^{30}B^2\>$ \\
\hline 
$6$   & $9.981077191$ & $14.51986635$ & $3.981253675$\\
$10$ & $\underline{5}.195580912$ & $7.113064471$ & $\underline{1}.811209013$\\
$14$ & $\underline{5.07}4025608$ & $\underline{6.93}2959071$ & $\underline{1.75}3605763$\\
$18$ & $\underline{5.0762}67533$ & $\underline{6.9362}45284$ & $\underline{1.754}098130$\\
$22$ & $5.076296208$ & $6.936287877$ & $1.754105280$\\
\hline\hline
\end{tabular}
\caption{The two-length trajectory bootstrap results for the independent Green's functions of length-$30, 32$. 
In spite of the significant deviations at $L_\text{max}=6$, 
these relatively long Green's functions converge rather rapidly as $L_\text{max}$ increases. 
The results for $\<\text{Tr} A^{32}\>$ are consistent with and more accurate than the Monte Carlo result $6.96(8)$ in \cite{Jha:2021exo}. 
 }
\label{table-2mm-5}
\end{table}

We can estimate the errors without resorting to the positivity bounds by 
examining how the predictions change as $L_{\text{max}}$ increases. 
The stable digits are interpreted as the reliable predictions. 
For example, the determinations of  
$\<\text{Tr}A^2\>$ are $(0.4, 0.42,0.4217,0.421784)$ for $L_{\text{max}}=6,10,14,18$. 
As the number of stable digits increases with $L_{\text{max}}$, 
the errors decrease with $L_{\text{max}}$, 
which are compatible with the positivity  bounds. 
In table \ref{table-2mm}, the convergence pattern further suggests that the accuracy of $(z_\ast, \<\text{Tr}A^2\>, \<\text{Tr}A^4\>, \<\text{Tr}A^2B^2\>)$ is around 8 digits for $L_\text{max}=22$. 

Above,  
we do not present the results for the coefficients $b_{k_1,k_2}$ of the analytic ansatz \eqref{2mm-analytic-ansatz-G}.  
The reason is that they converge much slower than the largest eigenvalues and the Green's functions. 
For example, the results for the low order coefficients at $L_\text{max}=10,14,18,22$ are  
$b_{0,0}\approx (0.9677, 1.0287,1.0134, 1.0149)$ and 
$b_{0,1}\approx (0.450,-0.183,0.296,0.282)$. 
\footnote{The coefficient $b_{0,2}$ seems to approach zero. } 
The slow convergence rate of these expansion coefficients may be related to 
the uniform convergence of the eigenvalue density on the branch cut 
(see figure \ref{fig:2mm-rho-1-error}), 
which is different from the local convergence near the edges of the eigenvalue support. 
Therefore, this is also consistent with the rapid convergence of the Green's functions. 


\subsection{Eigenvalue distributions}
\label{sec:eigenvalue-distributions}
Our conjecture \eqref{2mm-R-k-sum} about the singularity structure of the generating functions 
is closely related the distributions of eigenvalues. 
According to the contour integration formula \eqref{2mm-contour-integral},  
we can express the one-length and  two-length Green's functions as
\be
\<\text{Tr}\,A^{n}\>
=\int_{-z_\ast}^{z_\ast} dz\,z^{n}\, \rho^{(1)}(z)\,,\quad
\<\text{Tr}\,A^{n_1}B^{n_2}\>
=\int_{-z_\ast}^{z_\ast}  dz_1\int_{-z_\ast}^{z_\ast} dz_2\,z_1^{n_1}\,z_2^{n_2}\, \rho^{(2)}(z_1,z_2)\,,
\qquad
\label{2mm-rho-integral}
\ee
which are over the support of the eigenvalue distributions on the real axes. 
The one-length eigenvalue density $\rho^{(1)}(z)$ is related to the two-length eigenvalue density $\rho^{(2)}(z_1,z_2)$ by
\be
\rho^{(1)}(z_1)=\int_{-z_\ast}^{z_\ast} dz_2\,\rho^{(2)}(z_1,z_2)\,.
\label{rho1-rho2}
\ee
As the contours in \eqref{2mm-contour-integral}  extract 
the discontinuities on the branch cuts of the generating function $R^{(2)}(z_1,z_2)$. 
The two-length eigenvalue density is given by
\be
\rho^{(2)}(z_1,z_2)=-\frac {1}{\pi^2}
\sum_{k_1+k_2\leq k_\text{max}}\,b_{k_1,k_2}\,\frac{
(z_1^2-z_\ast^2)^{\frac 1 2+k_1}(z_2^2-z_\ast^2)^{\frac 1 2+k_2}}
{\frac{(-1/2-k_1)_{k_1+1}}{(k_1+1)!}\frac{(-1/2-k_2)_{k_2+1}}{(k_2+1)!}\,z_\ast^{2(k_1+k_2+2)}}\,
\,.
\label{2mm-rho-expansion}
\ee
\begin{figure}[h]
	\centering
		\includegraphics[width=0.8\linewidth]{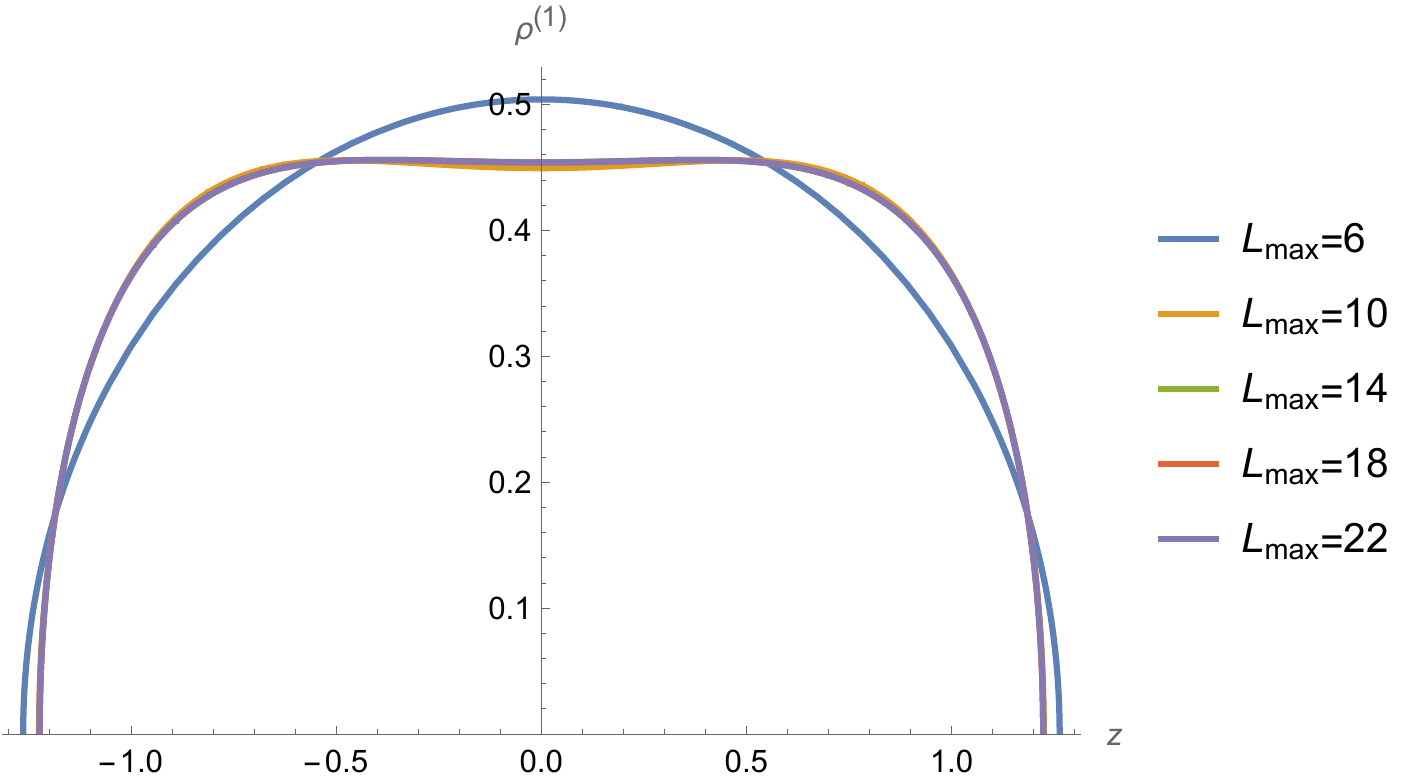}
	\caption{The one-length eigenvalue density $\rho^{(1)}(z)$ of the two-matrix model \eqref{2mm-definition} with $g=h=1$,  
	which encodes the information of the one-length Green's functions $\<\text{Tr}A^{n}\>$. 
	We use \eqref{rho1-rho2}, \eqref{2mm-rho-expansion} and the one-cut solutions at 
	$L_\text{max}=6,10,14,18,22$ to evaluate $\rho^{(1)}(z)$. 
	As our results converge rapidly, the curves for $L_\text{max}=14,18,22$ are indistinguishable from each other. 
	The eigenvalue distribution is consistent with and more accurate than the Monte Carlo results in  \cite{Jha:2021exo}.  }
	\label{fig:2mm-rho-1}
\end{figure}
\begin{figure}[h]
	\centering
		\includegraphics[width=0.8\linewidth]{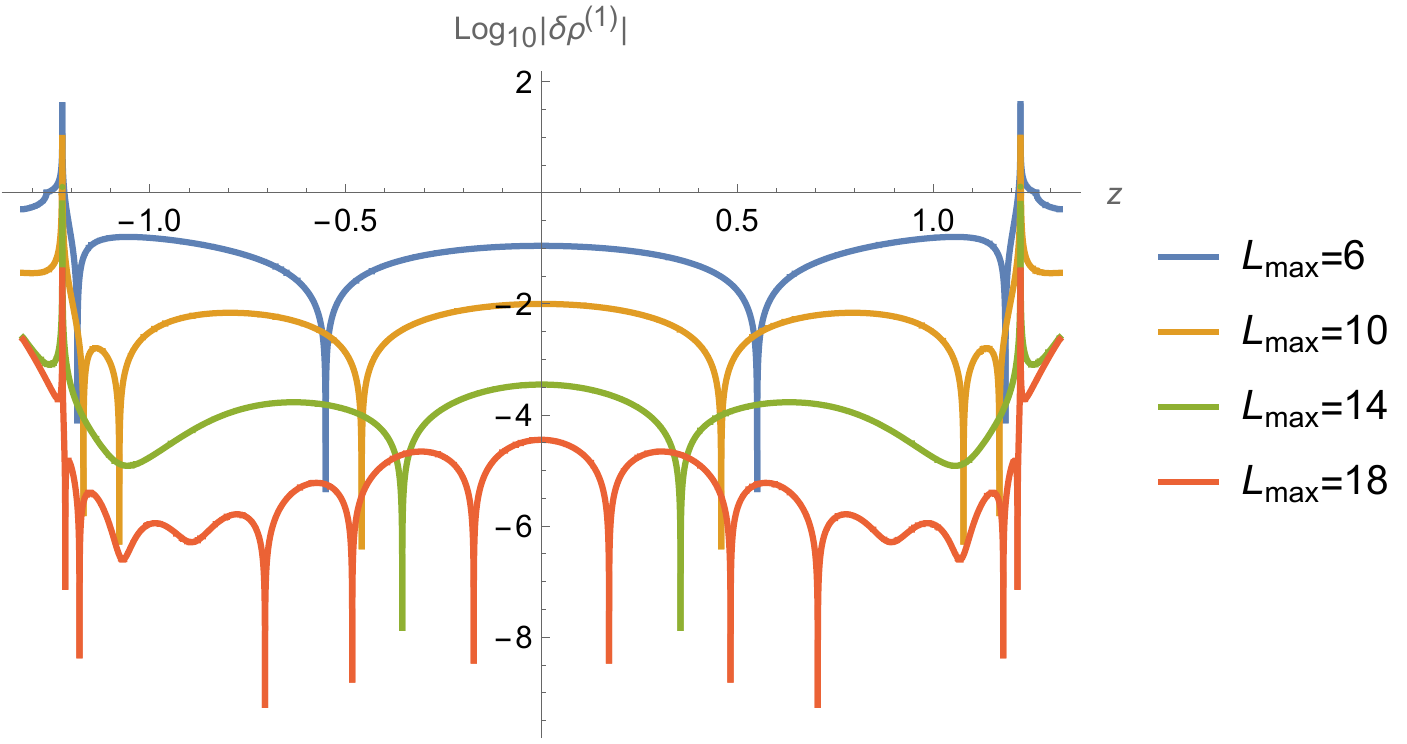}
	\caption{The relative error of the eigenvalue density $\d\rho^{(1)}=\rho^{(1)}/\rho^{(1)}_\text{ref}-1$ for various $L_\text{max}$. 
	We choose the $L_\text{max}=22$ solution as the reference density $\rho^{(1)}_\text{ref}$. 
	As $L_\text{max}$ increases, the approximate density improves uniformly on the support of the eigenvalue distribution. }
	\label{fig:2mm-rho-1-error}
\end{figure}
\begin{figure}[h]
	\centering
		\includegraphics[width=0.45\linewidth]{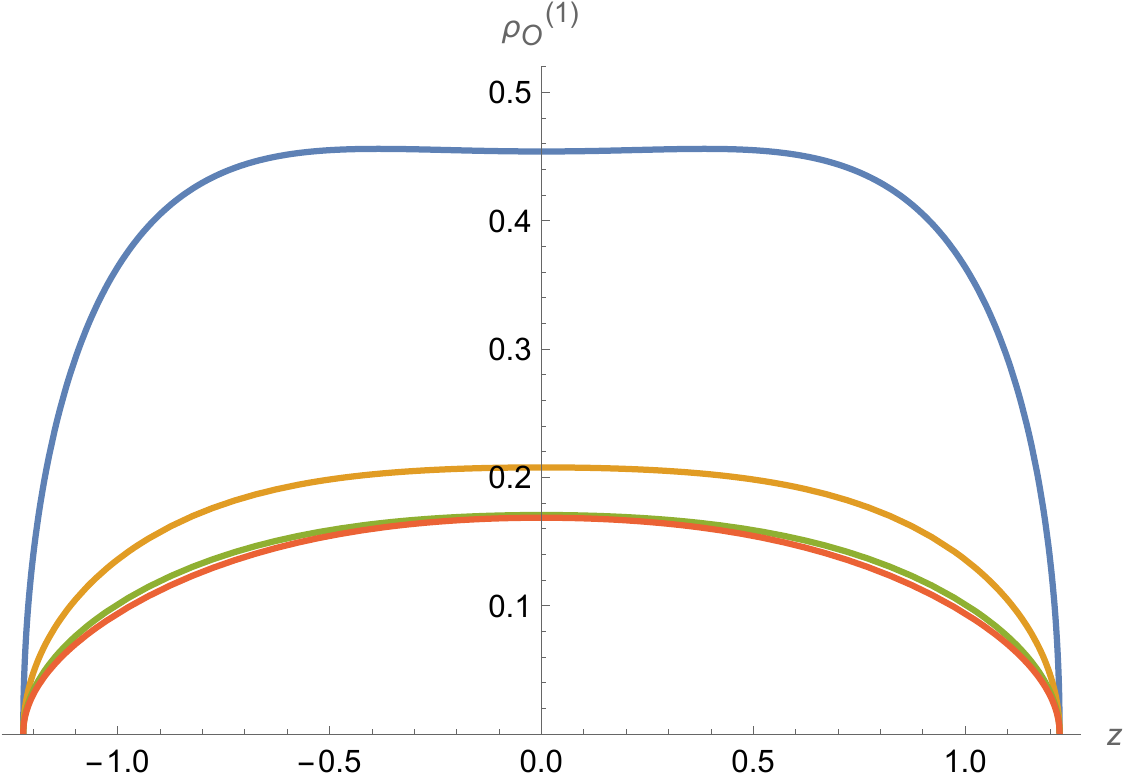}
		\includegraphics[width=0.45\linewidth]{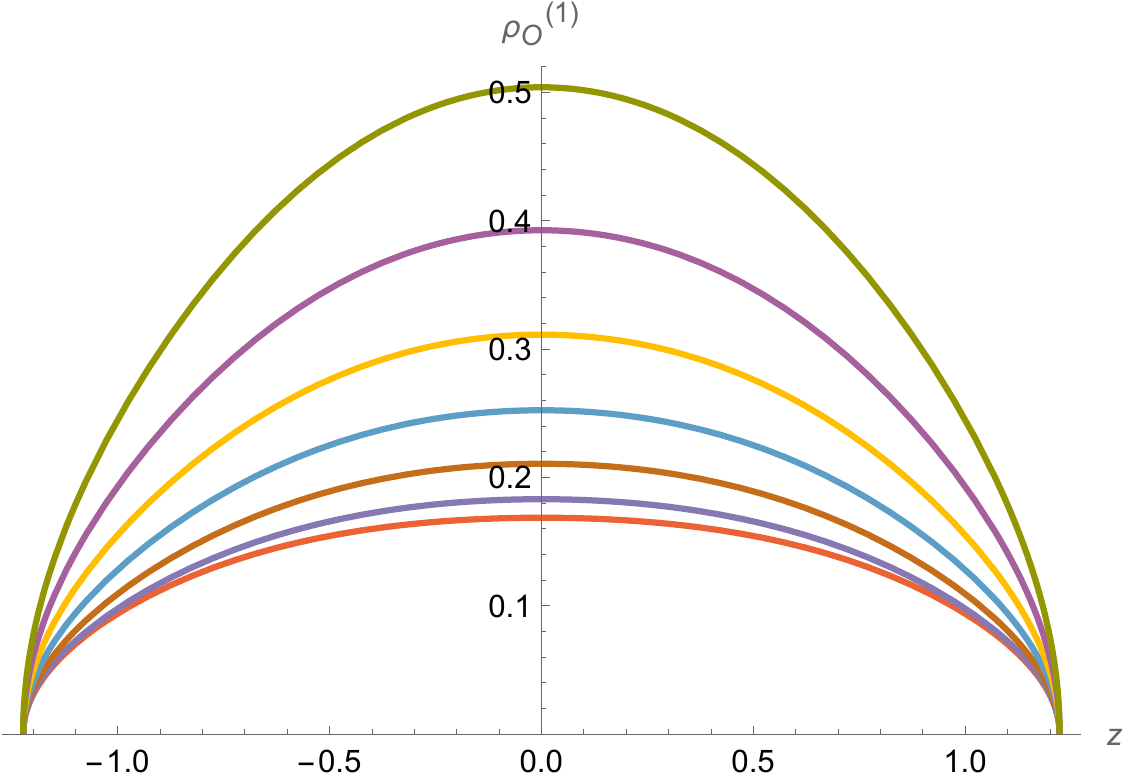}
		\includegraphics[width=0.07\linewidth]{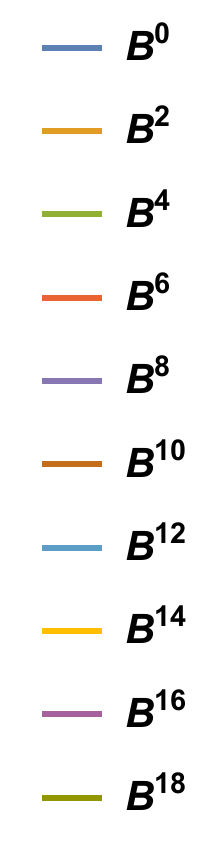}
	\caption{The generalized one-length eigenvalue densities $\rho_{\mathcal O}^{(1)}(z)$ of the two-matrix model \eqref{2mm-definition} with $g=h=1$, 
	which are associated with the generalized one-length Green's functions $\<\text{Tr}\,A^{n}\mathcal O\>$, 
	where $\mathcal O=B^0, B^2, \dots, B^{18}$. 
	We evaluate them using \eqref{2mm-rho-expansion}, \eqref{rho1-rho2-generalized} and the one-cut solution at $L_\text{max}=22$. }
	\label{fig:2mm-rho-1-generalized}
\end{figure}
\begin{figure}[h]
	\centering
		\includegraphics[width=0.8\linewidth]{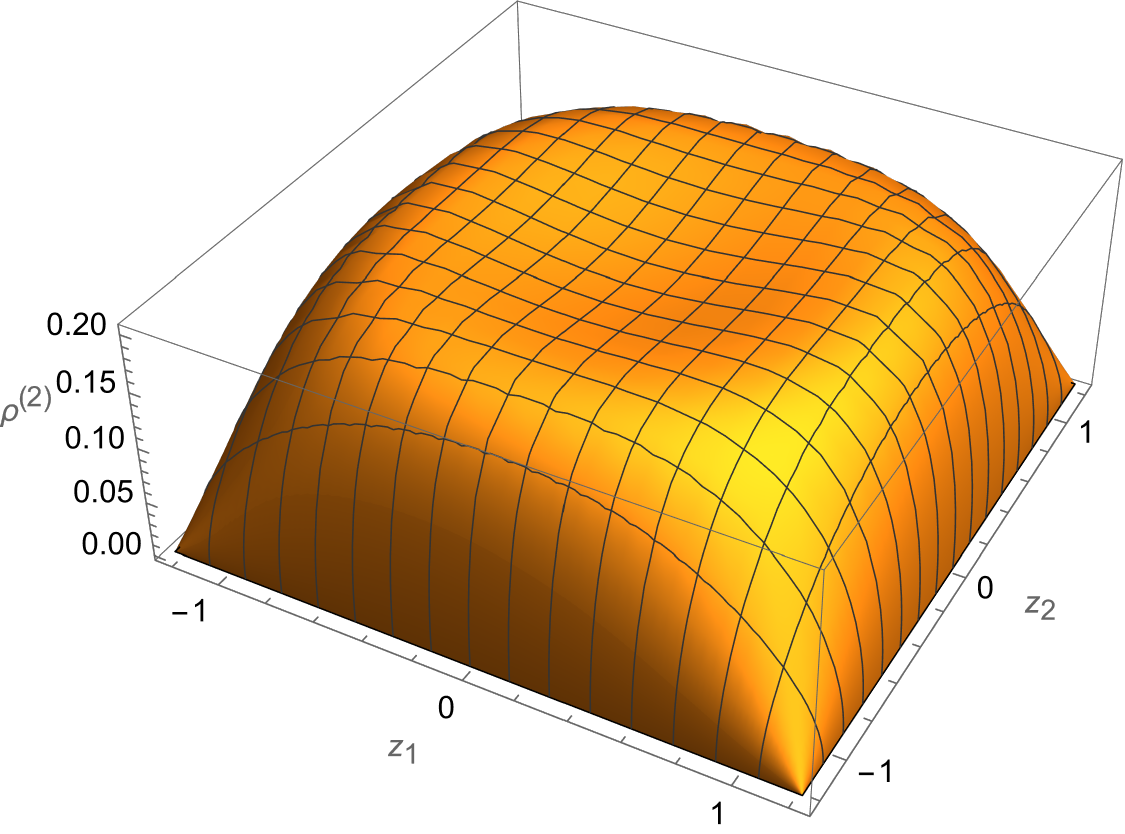}
	\caption{The two-length eigenvalue density $\rho^{(2)}(z_1,z_2)$ of the two-matrix model \eqref{2mm-definition} with $g=h=1$. 
	The density of eigenvalues is related to the two-length Green's functions $\<\text{Tr}A^{n_1}B^{n_2}\>$. 
	We use \eqref{2mm-rho-expansion} and the 1-cut solution at $L_\text{max}=22$ to evaluate $\rho^{(2)}(z_1,z_2)$.   }
	\label{fig:2mm-rho-2}
\end{figure}

In figure \ref{fig:2mm-rho-1}, we present the eigenvalue density $\rho^{(1)}(z)$ 
associated with the one-length Green's functions $\<\text{Tr}\,A^n\>$ for the length cutoff $L_{\text{max}}=6,10,14,18,22$, 
which converges rapidly as $L_{\text{max}}$ increases. 
Our results for the eigenvalue distribution confirm and are more accurate than the Monte Carlo results in \cite{Jha:2021exo}. 
In figure \ref{fig:2mm-rho-1-error}, we further show that the convergence of $\rho^{(1)}(z)$ is uniform on 
the support of the eigenvalue distribution. 
\footnote{Our ansatz can be viewed as rational approximations of the eigenvalue distributions. }
Using the two-length eigenvalue density $\rho^{(2)}(z_1,z_2)$, 
we can also compute the generalized one-length eigenvalue densities,  
such as $\rho_{B^2}^{(1)}(z)$, $\rho_{B^4}^{(1)}(z)$, $\rho_{B^6}^{(1)}(z)$.  
They are associated with the generalized resolvent $R_{\mathcal O}(z)$ in \eqref{generalized-resolvent} 
and the one-length Green's function 
\be
\<\text{Tr}\,A^n\mathcal O\>=\int_{-z_\ast}^{z_\ast} dz\,z^{n}\, \rho_{\mathcal O}^{(1)}(z)\,.
\label{2mm-rho-integral-generalized}
\ee
where $\rho_{\mathcal O}^{(1)}(z)$ with $\mathcal O=B^2, B^4, B^6, \dots$ are given by
\be
\rho_{B^{n_2}}^{(1)}(z_1)=\int_{-z_\ast}^{z_\ast} dz_2\,z_2^{n_2}\,\rho^{(2)}(z_1,z_2)\,.
\label{rho1-rho2-generalized}
\ee
The standard one-length eigenvalue density in \eqref{rho1-rho2} corresponds to the case of $\mathcal O=B^0$. 
In figure \ref{fig:2mm-rho-1-generalized}, we present the results for $\rho_{\mathcal O}^{(1)}(z)$ with $\mathcal O=B^0,B^2,\dots, B^{18}$. 
As the even integer $n$ grows, we find that $\rho_{B^{n}}^{(1)}(0)$ decreases for $n\leq 6$ 
and increases for $n\geq 6$. 

The two-length eigenvalue density does not take a simple factorized form due to the nontrivial angle dependence in the matrix integrals. 
In figure \ref{fig:2mm-rho-2},
we present the two-length eigenvalue density $\rho^{(2)}(z_1,z_2)$ 
associated with the two-length Green's functions $\<\text{Tr}\,A^{n_1}B^{n_2}\>$ for $L_\text{max}=22$. 

As a consistency check, we verify that the integrals in \eqref{2mm-rho-integral} and 
\eqref{2mm-rho-integral-generalized} 
with the eigenvalue density \eqref{2mm-rho-expansion} 
reproduce the results for the independent Green's functions in table \ref{table-2mm}-\ref{table-2mm-5}.

\section{Symmetry breaking solutions} 
\label{sec:symmetry-breaking}
Above, we assume the solutions of the two-matrix model are invariant under the $\mathbb Z_2$ transformations
\be
A\leftrightarrow -A\,,\quad
B\leftrightarrow -B\,, 
\ee
so the Green's functions with an odd number of $A$ or $B$ should vanish. 
If the coefficients of the quadratic terms are negative, 
we have a double well potential. 
It is natural to consider symmetry breaking solutions,  
as the tunneling effect is suppressed by the thermodynamic-like large-$N$ limit. 
To capture the symmetry breaking phenomena, 
we need to extend the symmetric one-cut ansatz to more complicated forms. 

In section \ref{sec:1mm-breaking}, we will briefly review the symmetry breaking solutions in the one-matrix model, 
which provides natural building blocks for the two-matrix ansatz. 
In section \ref{sec:2mm-breaking}, 
we will use more complicated ansatzes to study the solutions 
with spontaneously broken symmetries in the two-matrix model. 

\subsection{One-matrix model}
\label{sec:1mm-breaking}
Let us consider the potential with a negative coefficent for the quadratic term
\be
V(M)=-\frac 1 2 M^2+\frac g 4 M^4\,,
\label{1mm-potential-negative}
\ee
where $g>0$. 
We refer to \cite{Lin:2020mme,Kazakov:2021lel} for discussions 
with more emphasis on the positivity constraints. 
Some main features of this one-matrix model extend to the two-matrix version with $\text{tr}[A,B]^2$ interaction. 

As in \eqref{1mm-1cut-resolvent}, we can consider the one-cut solutions. 
If we do not impose the $\mathbb Z_2$ symmetry, 
the one-cut ansatz for the resolvent reads
\be
R^{(\text{one-cut})}(z)=\sum_{k=0}^{k_\text{max}} a_k\, z^k\, \sqrt{(z-z_{\ast,1})(z-z_{\ast,2})}+\text{regular}\,,
\label{1mm-asymmetric-one-cut-resolvent}
\ee
where $k_\text{max}=2$ for the quartic potential \eqref{1mm-potential-negative}. 
As in \eqref{1mm-buliding-block-large-z} and \eqref{1mm-Gn-ansatz}, 
we can extract the single trace moments from the large $z$ expansion of the singular part of the resolvent. 
Then we can determine the free parameters by the loop equations
\be
-G_{n+1}+gG_{n+3}=\sum_{p=0}^{n-1}G_pG_{n-p-1}\,.
\ee
For example, we obtain $a_2=-g/2$. 
We have two types of one-cut solutions:
\begin{itemize}
\item
$\<\text{Tr}M\>=0$

In this case, the single trace moment $\<\text{Tr}M^{n}\>$ vanishes if $n$ is an odd number, 
so the $\mathbb Z_2$ symmetry is preserved. 
As the loop equations require that $a_1=0$ and $z_1=-z_2$, 
the solution takes the form in \eqref{1mm-1cut-resolvent}. 
The other parameters can be expressed in terms of $z_\ast^2=z_1^2=z_2^2$, such as
\be
\<\text{Tr} M^2\>=\left(\frac 1 {3}+\frac{1}{36g}\right)z^2_\ast+\frac 1 {9g}\,,
\quad
a_0=\frac {(z^2_\ast)^2}{512}-\left(\frac g 4+\frac 1 {384 g}\right)z^2_\ast+\frac 1 2-\frac 1 {96g}\,.
\ee
Since $z_\ast^2$ satisfies the quadratic equation
$3g(z_\ast^2)^2-4z_\ast^2-16=0$, 
we have two solusions
\be
z^2_{\ast,\pm}=\frac {2} {3g}\left(1\pm \sqrt{1+12g}\right)\,.
\label{1mm-dw-1-cut-root}
\ee
The branch points associated with the $z^2_{\ast,-}$ solution are on the imaginary axis. 
Since $z^2_{\ast,+}$ is positive, the corresponding eigenvalue distribution is supported on the real axis.   
The eigenvalue density of the $z^2_{\ast,+}$ solution is positive on the support for $g\geq 1/4$, 
but it becomes negative around $z=0$ for $g<1/4$. 
\footnote{We thank Zechuan Zheng for clarifying this point. }

\item
$|\<\text{Tr}M\>|>0$

In this case, the $\mathbb Z_2$ symmetry is spontaneously broken. 
The symmetry breaking solution can be parametrized by the order parameter $\<\text{Tr}M\>$. 
For example, we have
\be
\<\text{Tr}M^2\>=\frac{1}{3g(1+60g)}\left(3+52g+125g^2\,\<\text{Tr}M\>^2\right)\,,
\ee
which is positive for the real solutions of $\<\text{Tr}M\>$. 
One can show that $\<\text{Tr}M\>^2$ satisfies the quadratic equation
\be
28125g^4x^2-8g(2+75g+4500g^2)x+16(1+12g)^3=0\,.
\ee
The explicit one-cut solutions for $\<\text{Tr}M\>_{ \pm}^2$ are
\be
\<\text{Tr}M\>_{ \pm}^2=\frac{4}{28125g^3}
\left(2+75g+4500g^2\pm2(1+60g)(1-15g)^{3/2}\right)\,,
\label{1mm-asym-1-cut-solution}
\ee
which are positive real for $0<g\leq \frac 1 {15}$. 
For $g>\frac 1 {15}$, 
the one-cut solutions in \eqref{1mm-asym-1-cut-solution} are a pair of complex conjugates.  
As $g$ decreases, they meet at $g=\frac 1 {15}$ and then become real solutions for $0<g\leq \frac 1 {15}$. 
Although the corresponding branch points of the resolvents are all located on the real axis, 
those of the $\<\text{Tr}M\>_{+}$ solution have opposite signs for $0<g<\frac 1 {20}$. 
Furthermore,  the eigenvalue density of the $\<\text{Tr}M\>_{+}$ solution is negative at small $|z|$, 
while that of $\<\text{Tr}M\>_{-}$ solution is always positive.  
Therefore, $\<\text{Tr}M\>_{-}$ is more physical than $\<\text{Tr}M\>_{+}$ 
in the range $0<g\leq \frac 1 {15}$. 
\end{itemize}

For $0<g<1/4$, we should introduce a more complicated ansatz  
in order to capture the generic $\mathbb Z_2$ symmetry breaking behavior. 
We consider the next-to-minimally singular solutions, 
i.e. the two-cut solutions.
The corresponding ansatz for the resolvent reads
\be
R^{(\text{two-cut})}(z)=\sum_{k=0}^{k_\text{max}} a_k\, z^k\sqrt{(z-z_{\ast,1})(z-z_{\ast,2})(z-z_{\ast,3})(z-z_{\ast,4})}+\text{regular}\,,
\label{1mm-2-cut-resolvent}
\ee
where $k_\text{max}=1$ for the quartic potential \eqref{1mm-potential-negative}. 
The two-cut solutions can be parametrized by the symmetry breaking order parameter $\<\text{Tr}M\>$. 
The more physical solutions have eigenvalues supported on the real axis, 
then we assume $z_{\ast,1}\geq z_{\ast,2}\geq z_{\ast,3}\geq z_{\ast,4}$. 
Although the building blocks of \eqref{1mm-2-cut-resolvent}  do not lead to simple analytic expressions, 
we can still extract the single trace moments 
from the large $z$ expansion of the singular part, 
as in \eqref{1mm-buliding-block-large-z} and \eqref{1mm-Gn-ansatz}. 
Then we can use the loop equations to 
solve for the free parameters, such as $a_1=-g/2$. 

Let us first examine the $\mathbb Z_2$ symmetric limit of the two-cut solutions
\be
z_{\ast,1}=-z_{\ast,4}\,,\quad z_{\ast,2}=-z_{\ast,3}\,,\quad 
a_0=0\,.
\ee 
In this special case, the eigenvalues are symmetrically distributed near the two minima of the effective potential for the matrix eigenvalues. 
The $\mathbb Z_2$ symmetric two-cut resolvent takes a simple form
\be
R^{\text{(two-cut, s)}}(z)=a_1\, z\sqrt{(z^2-z_{\ast,1}^2)(z^2-z_{\ast,2}^2)}+\text{regular}\,,
\ee
which implies $\<\text{Tr}M\>=0$.
The remaining parameters can be determined by the loop equations. 
We have
\be
R^{\text{(two-cut, s)}}(z)=-\frac g 2\, z\,\sqrt{\left(z^2-\frac{1-2\sqrt g}{g}\right)\left(z^2-\frac{1+2\sqrt g}{g}\right)}+\text{regular}\,,
\label{1mm-symmetric-2-cut-solution}
\ee
which implies $\<\text{Tr} M^2\>=\frac 1 g$. 
For $0<g< 1/4$, the eigenvalue distribution is supported on the real axis with positive eigenvalue density. 
For $g=1/4$, two branch points approach zero and we obtain the symmetric one-cut solution associated with $z^2_{\ast,+}$ in \eqref{1mm-dw-1-cut-root}. 
For $g>1/4$, we have $1-2\sqrt g<0$, so the two branch points are a pair of complex conjugates on the imaginary axis.

\begin{figure}[h]
	\centering
		\includegraphics[width=0.8\linewidth]{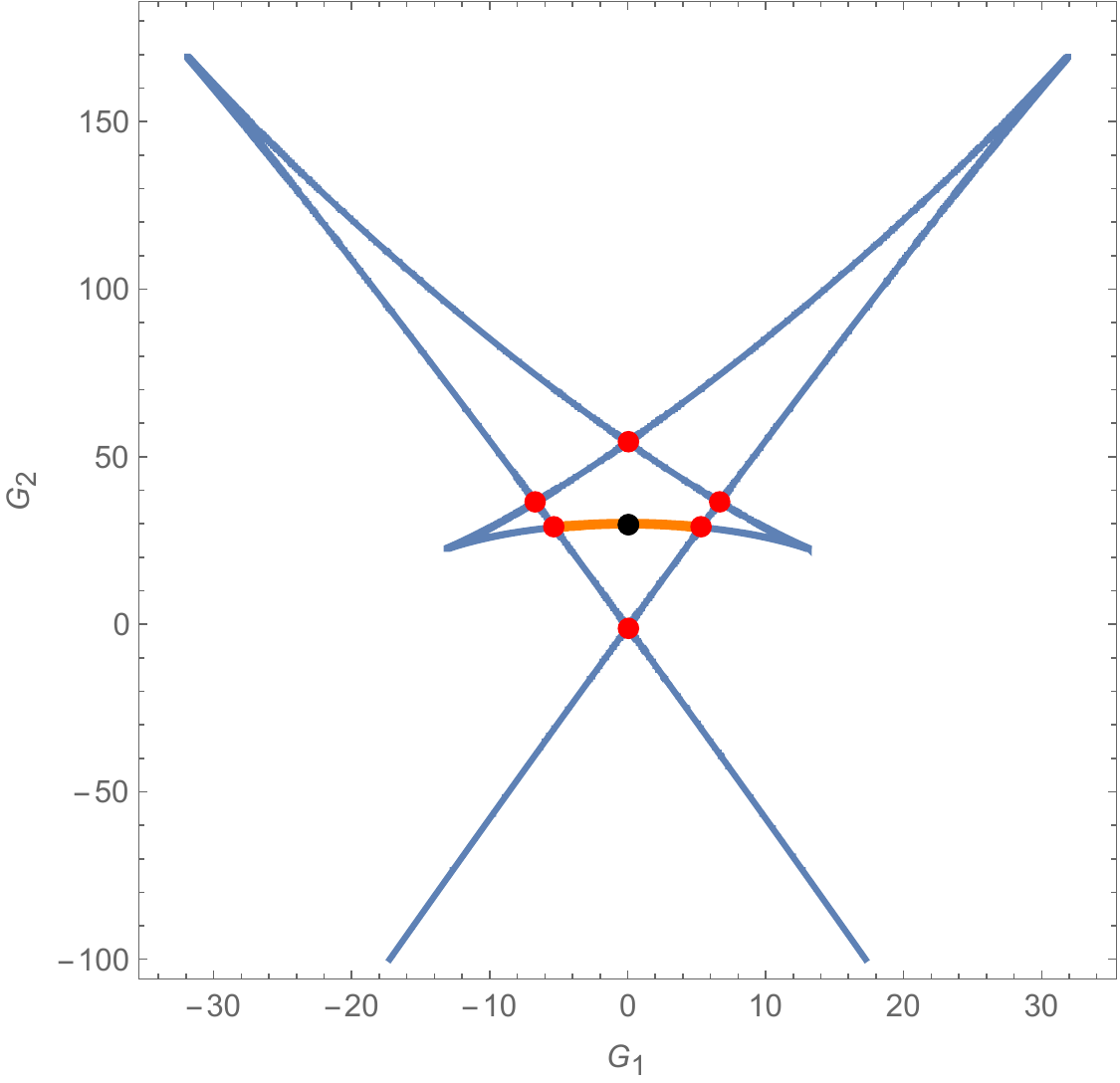}
	\caption{The two-cut solutions of the large-$N$ one-matrix model with the potential $V(M)=-\frac 1 2 M^2+\frac g 4 M^4$, where $g=\frac 1 {30}$, $G_1=\<\text{Tr}M\>$ and $G_2=\<\text{Tr}M^2\>$.  The red dots represent the one-cut solutions \eqref{1mm-dw-1-cut-root} and \eqref{1mm-asym-1-cut-solution}, 
	which are located at the intersection points of the two-cut solutions. 
	The black dot indicates the $\mathbb Z_2$ symmetric two-cut solution \eqref{1mm-symmetric-2-cut-solution}. 
	The two-cut solution with positive density of eigenvalues supported on the real axis is denoted by the orange curve, 
	whose endpoints are the one-cut solutions at 
	$(G_1, G_2)=(\pm\frac{4}{5}\sqrt{57-9\sqrt{2}},\frac 1 3(98-8\sqrt 2))\approx(\pm 5.32298, 28.8954)$. }
	\label{fig:1mm-2-cut-contour}
\end{figure}

After examining the special cases,  
we consider the generic two-cut solution in \eqref{1mm-2-cut-resolvent}.  
In the limit $\<\text{Tr} M\>\rightarrow 0$, 
the more physical branch of solution should reduce to  
the $\mathbb Z_2$ symmetric one-cut solution \eqref{1mm-dw-1-cut-root} for $g\geq 1/4$ 
and the $\mathbb Z_2$ symmetric two-cut solution \eqref{1mm-symmetric-2-cut-solution} for $0<g< 1/4$. 
For a given $\<\text{Tr} M\>$, we obtain a quintic equation for $\<\text{Tr} M^2\>$,
so there are five branches of solutions. 
The other parameters can be expressed in terms of $\<\text{Tr} M^2\>$. 
For $g=\frac 1 {30}$, the explicit algebraic equation for $G_1=\<\text{Tr} M\>$ and  $G_2=\<\text{Tr} M^2\>$ reads
\be
&&G_2^5- \frac{410}{3}\, G_2^4+\frac 5 9 \left(10696 - 75  G_1^2\right)G_2^3
+\frac{250}{27}\left(-8296 + 513 G_1^2\right)G_2^2
\nn
&&+\frac{25}{324} \left(-2085824 - 2566320 G_1^2 + 3375 G_1^4\right)G_2
\nn
&&+\frac{25}{1944} \left(-6094080 + 218680896 G_1^2 - 681900 G_1^4 + 625 G_1^6\right)=0\,.
\ee
The generic $g$ version of this algebraic equation can be found in \cite{Kazakov:2021lel}. 
In figure \ref{fig:1mm-2-cut-contour}, we present the real solutions of $(G_1, G_2)$ from the two-cut ansatz \eqref{1mm-2-cut-resolvent} with $g=\frac 1 {30}$. 
We notice that 
\begin{itemize}
\item
The one-cut solutions can be connected by two-cut solutions.
\item 
The one-cut solutions emerge at the intersection points of the two-cut solutions. 
\end{itemize}
As $G_1$ changes, some two-cut solutions for $G_2$ may collide and become a pair of complex conjugates, so they annihilate each other from the perspective of real solutions. 
We do not find any one-cut solutions associated with these annihilations. 
As we increase $g$, the two symmetry breaking one-cut solutions meet at $g=\frac 1 {15}$ and become a pair of complex conjugates for $g>\frac 1 {15}$. 
For $\frac 1 {15}<g<\frac 1 4$,  the more physical branch of real solution cannot reach the one-cut limit because the annihilations happen at smaller $|\text{Re}(G_1)|$. 
If we further increase $g$, the more physical branch of solution shrinks to 
a point at $g=\frac 1 4$. 
For $g\geq \frac 1 4$, the more physical solution is given by the $\mathbb Z_2$ symmetric one-cut solution 
associated with $z^2_{\ast,+}$ in \eqref{1mm-dw-1-cut-root}.

Using the analytic ansatzes for the singular part of the resolvent, 
we are able to study the various types of one-cut solutions and two-cut solutions. 
Although some solutions are more physical than the others, 
the intersections or collisions of the less physical solutions 
help us to understand better the more physical ones, 
such as their range and evolution behavior as the coupling constant $g$ changes. 
Below, we will apply this approach to the more complicated two matrix model with $\text{tr}[A,B]^2$ interaction. 

\subsection{Two-matrix model}
\label{sec:2mm-breaking}
To study the symmetry breaking solutions of the two-matrix model, 
we consider the following potential
\be
V(A,B)=-\frac 1 2 A^2+\frac g 4 A^4-\frac 1 2 B^2+\frac g 4 B^4-\frac h 2 [A,B]^2\,.
\label{2mm-potential-negative}
\ee
For $h=0$, we have two decoupled one-matrix models, 
which allow for symmetric breaking solutions for $0<g<\frac 1 4$.
Below, we focus on the case of $g=\frac 1 {30}, h=\frac 1 {15}$, 
which is not solvable by the known analytic methods.
If the $\mathbb Z_2$ symmetry is spontaneously broken, 
there are more independent Green's functions. 
For $L_\text{max}< 16$, we can again choose the two-length Green's functions as the independent ones, 
such as those in \eqref{2mm-independent-G} and 
\be
G_{0,1}\,,\quad
G_{1,1}\,,\quad
G_{1,2}\,,\quad
G_{0,5}\,,\quad
G_{0,7}\,,\quad
G_{1,7}\,,\quad
G_{1,11}\,,\quad
G_{0,13}\,,\quad
G_{1,14}\,.
\ee
We assume that the $\mathbb Z_2$ symmetry $A\leftrightarrow B$ is unbroken, 
so we have $G_{n_1,n_2}=G_{n_2,n_1}$.
We also use the simplified notation $G_{n}=G_{0,n}$. 
Below we mainly consider $L_\text{max}\leq 8$. 
The loop equations imply some self-consistent constraints for the two-length Green's functions. 
Some explicit examples are
\be
G_{0,3}=-30G_{0,1}\,,\quad
G_{1,3}=G_{1,1}\,,\quad G_{2,3}=30G_{1,2}\,,\label{breaking-constraint-1}
\ee
\be
G_{1,4}=-450G_{0,1}+30G_{1,2}+\frac {1}{2}G_{0,5}\,,\label{breaking-constraint-2}
\ee
\be
G_{2,4}=\frac{5 }{2}G_{0,1}^2-70 G_{0,2}+30 G_{2,2}-5G_{0,4}+\frac{1}{4}G_{0,6}-75\,,
\label{breaking-constraint-3}
\ee
\be
G_{1,5}=G_{0,2} G_{1,1}+G_{0,1}G_{1,2}+900 G_{1,1}\,,\quad
G_{3,3}=900G_{1,1}\,,\label{breaking-constraint-4}
\ee
\be
G_{4,4}&=&\frac{1}{11}\Big(600 G_{0,1}^2+80G_{1,1}^2+175 G_{0,2}^2-38250G_{0,2}+15865 G_{2,2}-5360G_{0,4}+190G_{0,6}
\nn&&\qquad+G_{0,8}-36000\Big)+25 G_{1,2} G_{0,1}-\frac{13G_{2,6}}{22}\,.
\label{breaking-constraint-5}
\ee
For symmetric solutions, the nontrivial constraints are associated with \eqref{breaking-constraint-3} and \eqref{breaking-constraint-5}. 
We do not present the length-7 and other length-8 constraints as they are not used below.
As in the case of the one-matrix model, 
the symmetry breaking solutions of the two-cut type can be parametrized by the order parameter
\be
G_1=\<\text{Tr}A\>=\<\text{Tr}B\>\,.
\ee
As a generalization of the \eqref{1mm-2-cut-resolvent}, 
we introduce to a two-cut ansatz for the two-length generating function
\be
R^{(2)}(z_1,z_2)&=&\sum_{k_1=0}^{k_\text{max}}\,\sum_{k_2=0}^{k_\text{max}-k_1}\,a_{k_1,k_2}
\prod_{j=1,2}\, z_j^{k_j}\,\sqrt{(z_j-z_{\ast,1})(z_j-z_{\ast,2})(z_j-z_{\ast,3})(z_j-z_{\ast,4})}
\nn&&+\text{regular}\,.
\label{2mm-breaking-2-cut-ansatz}
\ee
The support for the eigenvalue distribution is encoded in the locations of the branch points. 
There are 4 different branch points in \eqref{2mm-breaking-2-cut-ansatz}. 

In the one-cut limit, two branch points meet $z_{\ast,3}-z_{\ast,4}\rightarrow 0$ 
and become a regular point located at $z_{\ast,3}$. 
Accordingly, the one-cut ansatz for the two-length generating function reads
\be
R^{(2)}(z_1,z_2)&=&\sum_{k_1=0}^{k_\text{max}}\,\sum_{k_2=0}^{k_\text{max}-k_1}\,a_{k_1,k_2}\, \prod_{j=1,2}\, z_j^{k_j}\,(z_j-z_{\ast,3})\sqrt{(z_j-z_{\ast,1})(z_j-z_{\ast,2})}
+\text{regular}\,,
\nn
\ee
where $(z_{\ast,1},z_{\ast,2})$ denote the locations of the two branch points. 
For $g=\frac 1 {30}$, 
we expect that $G_1\neq 0$ for the more physical solutions as in the case of the one-matrix model. 
For $k_\text{max}=0$, we have 4 parameters $(z_{\ast,1},z_{\ast,2},z_{\ast,3},a_{0,0})$, 
so we need 4 equations. The first equation is again the normalization condition $G_{0,0}=1$. 
If we impose the first constraint in \eqref{breaking-constraint-1}, then the second and third ones in \eqref{breaking-constraint-1} are automatically satisfied. 
Then we can make use of the two constraints in \eqref{breaking-constraint-2} and \eqref{breaking-constraint-3}. 
We obtain 30 solutions and 20 of them are real solutions. 
For the positivity of the eigenvalue density, we further impose that $(z_{\ast,3}-z_{\ast,1})(z_{\ast,3}-z_{\ast,2})>0$, which selects two pairs of solutions. 
If the eigenvalues are distributed around a local minimum of the effective potential, 
the solution should satisfy $z_{\ast,1}z_{\ast,2}>0$. In the end, we find only one pair of solutions
\be
(G_1\,,\,G_2\,,\,z_{\ast,1}\,,\,z_{\ast,2})\approx(\pm5.3293\,,\, 28.9401\,,\, \pm6.6627\,,\,\pm3.6454)\,,
\label{asymmetric-one-cut-solution-1}
\ee
with $z_{\ast,3}\approx \pm1.9079$ and $a_{0,0}\approx 0.073275$. 
It is also natural to replace \eqref{breaking-constraint-2} with \eqref{breaking-constraint-5}, 
as the latter one does not vanish automatically for the symmetric solution. 
In this case, we have 56 solutions. 
After imposing the same additional requirements, we again obtain only one pair of solutions
\be
(G_1\,,\,G_2\,,\,z_{\ast,1}\,,\,z_{\ast,2})\approx(\pm5.3292, 28.9392\,,\,\pm6.6641\,,\,\pm3.6474)\,,
\label{asymmetric-one-cut-solution-2}
\ee
with $z_{\ast,3}\approx \pm 1.8770$ and $a_{0,0}\approx 0.071883$.
For $k_\text{max}=2$, there are 3 more free parameters $(a_{0,1}, a_{0,2}, a_{1,1})$, 
so we take into account the self-consistent constraints in \eqref{breaking-constraint-1}, \eqref{breaking-constraint-2}, \eqref{breaking-constraint-3}, together with the normalization condition $G_{0,0}=1$ 
and the first constraint in \eqref{breaking-constraint-4}. 
As the number of solutions increases significantly, 
we use the efficient package \texttt{HomotopyContinuation.jl} \cite{Breiding-Timme}
to solve the systems of polynomial equations here and below. 
Then we extract two pairs of solutions by the same requirements as above
\be
(G_1\,,\,G_2\,,\,z_{\ast,1}\,,\,z_{\ast,2})&\approx&(\pm 5.32892\,,\, 28.93747\,,\,\pm6.68471\,,\, \pm3.14985)\,,
\nn
&&(\pm 5.32897\,,\, 28.93779\,,\, \pm 6.68441\,,\,\pm3.58664)\,. 
\label{asymmetric-one-cut-solution-3}
\ee
For a different combination of the length-6 constraints, we do not find any real solution. 
It also natural to replace the first constraint in \eqref{breaking-constraint-4} with \eqref{breaking-constraint-5}, 
as the latter is nontrivial for the symmetric solution. 
Then the solutions are
\be
(G_1\,,\,G_2\,,\,z_{\ast,1}\,,\,z_{\ast,2})&\approx&(\pm 5.32878\,,\, 28.93663\,,\,\pm6.68555\,,\, \pm3.50902)\,,
\nn
&&
(\pm 5.32880\,,\, 28.93672\,,\, \pm6.68738\,,\,\pm3.18846)\,,
\nn
&&
(\pm 5.32891\,,\, 28.93741\,,\, \pm6.68819\,,\,\pm3.57615)\,. 
\label{1-cut-breaking-kmax-2}
\ee
As in the case of the one-matrix model with $g=\frac 1 {30}$, 
these one-cut solutions should be related to the endpoints of the more physical branch of symmetry breaking solutions. 
The estimates $(G_1,G_2)\approx (\pm 5.329, 28.937)$ are well consistent with 
the edges of the positivity bounds obtained for a much higher length cutoff $L_\text{max}=16$ in \cite{Kazakov:2021lel}. 

Another special limit is the $\mathbb Z_2$ symmetric solution
\be
z_{\ast,1}=-z_{\ast,4}\,,\quad z_{\ast,2}=-z_{\ast,3}\,,
\ee
and $a_{k_1,k_2}$ vanishes if $k_1$ or $k_2$ is an even integer. 
The lowest cutoff for the $\mathbb Z_2$ symmetric ansatz is $k_\text{max}=2$. 
The corresponding two-cut ansatz reads
\be
R^{(2)}(z_1,z_2)=a_{1,1}\,z_1\,z_2\,
\sqrt{(z^2_1-z^2_{\ast,1})(z^2_1-z^2_{\ast,2})}\sqrt{(z^2_2-z^2_{\ast,1})(z^2_2-z^2_{\ast,2})}
+\text{regular}\,.
\ee
which contains 3 free parameters $(a_{1,1}, z_{\ast,1}^2, z_{\ast,2}^2)$. 
They can be determined by the normalization condition $G_{0,0}=1$ 
and the length-6,8 constraints in \eqref{breaking-constraint-3}, \eqref{breaking-constraint-5}. 
We find 6 solutions, but only one of them is real
\be
(G_2\,,\, G_4\,,\, G_{2,2}\,,\, z^2_{\ast,1}\,,\, z^2_{\ast,2})
\approx
(29.7461\,,\, 902.788\,,\, 884.828\,,\, 6.1824^2\,,\, 4.6119^2)\,.
\label{2-cut-symmetric-solution-kmax-2}
\ee
The approximate solution for $G_2$ is also consistent with 
the upper bound from positivity constraints for $L_\text{max}=16$ in \cite{Kazakov:2021lel}.  

\begin{figure}[h]
	\centering
		\includegraphics[width=0.8\linewidth]{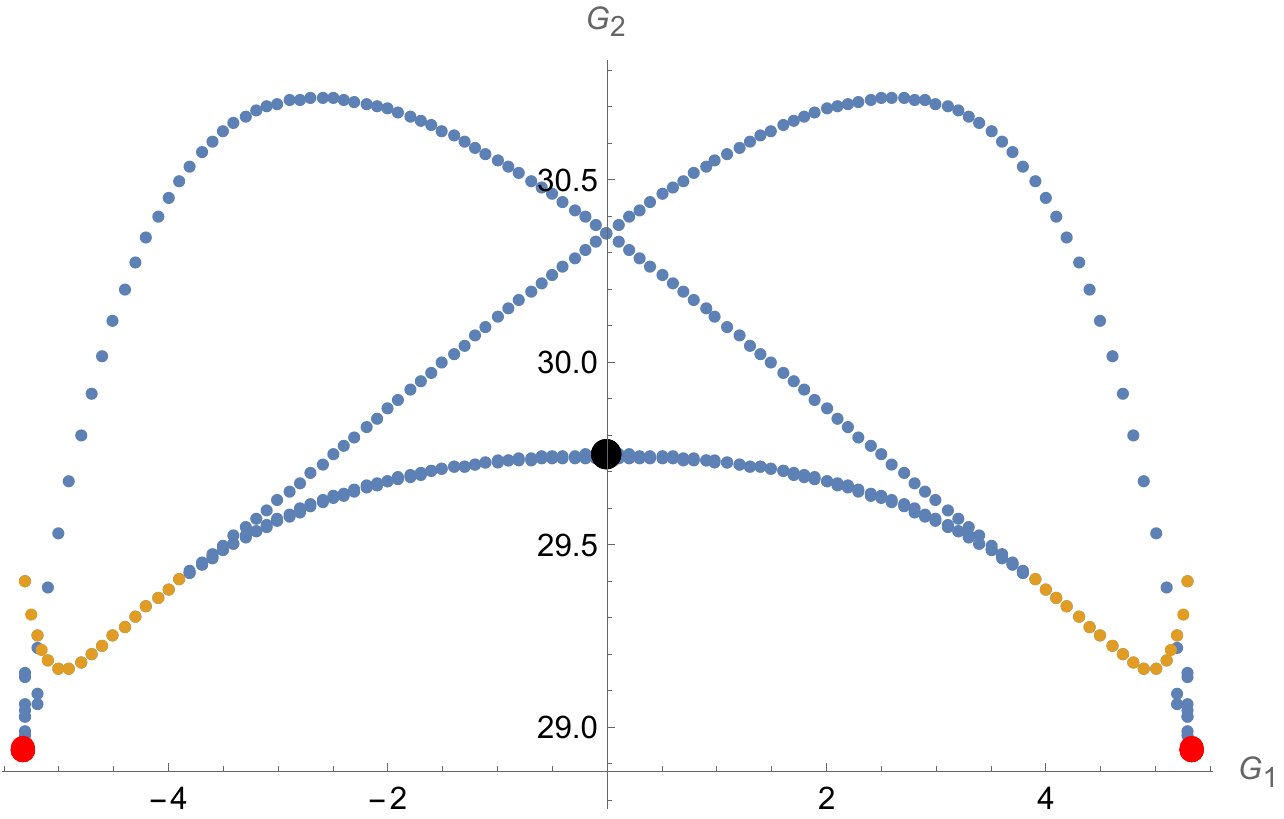}
	\caption{The two-cut solutions of the large-$N$ two-matrix model with the potential 
	$V(A,B)=-\frac 1 2 A^2+\frac g 4 A^4-\frac 1 2 B^2+\frac g 4 B^4-\frac h 2 [A,B]^2$, 
	where $g=\frac 1 {30}, h=\frac 1 {15}$, $G_1=\<\text{Tr}A\>$ and $G_2=\<\text{Tr}A^2\>$. 
	We use \eqref{2mm-breaking-2-cut-ansatz} with $k_\text{max}=2$ as the two-cut ansatz for the two-length generating function. We consider $G_1=\pm \frac{k}{10}$ with $k=10^{-5}$ and $k=1, 2, \dots, 52,53$. For $|G_1|\leq 3.8$, the lowest branch is associated with two nearly degenerate solutions. 
	The one connected to the symmetric two-cut solution annihilates 
	with the middle branch at $G_1\approx \pm 3.9$. 
	For $|G_1|\geq 3.9$, the lower branch and its continuation is represented by the orange dots. 
	The pairs of asymmetric one-cut solutions in \eqref{asymmetric-one-cut-solution-1}, \eqref{asymmetric-one-cut-solution-2}, \eqref{asymmetric-one-cut-solution-3}, \eqref{1-cut-breaking-kmax-2} are denoted by red dots, but they are not distinguishable. The symmetric two-cut solution in \eqref{2-cut-symmetric-solution-kmax-2} is indicated by the black dot. 
	 }
	\label{fig:2mm-2-cut-contour}
\end{figure}

\begin{figure}[h]
	\centering
		\includegraphics[width=0.8\linewidth]{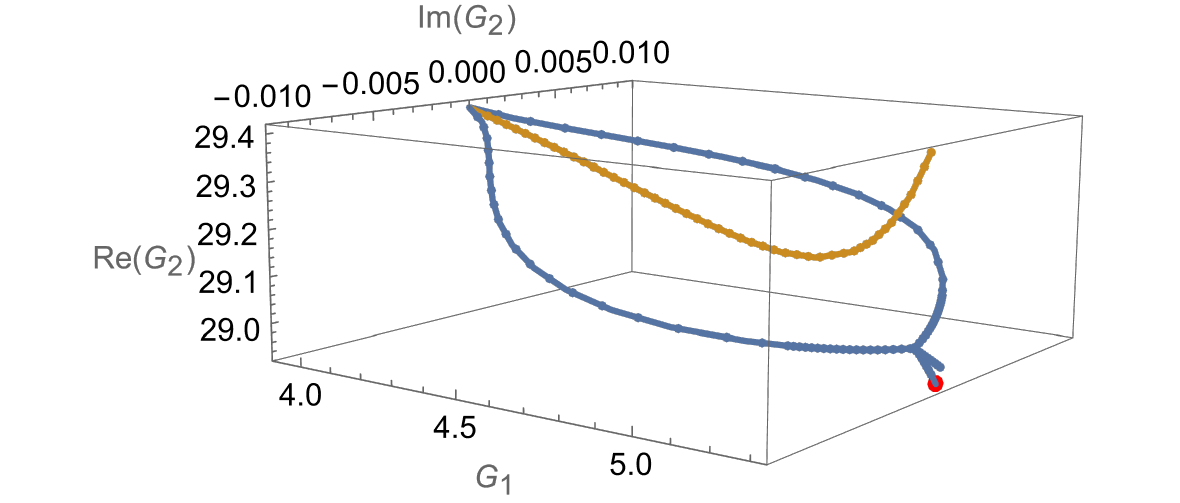}
	\caption{Some two-cut solutions of the two-matrix model in the complex space. 
	The real solution in orange is the same as the one in figure \ref{fig:2mm-2-cut-contour}. 
	The two solutions in blue meet at $|G_1|\approx3.9$ and become a pair of complex conjugates as $|G_1|$ increases. 
	They meet again at around $|G_1|=5.27$ and become real solutions. 
	At $|G_1|\approx 5.33$, the lowest branch is close to the asymmetric one-cut solutions 
	denoted by nearly degenerate red dots. }
	\label{fig:2mm-2-cut-contour-complex}
\end{figure}

After studying two special limits, 
we consider the generic symmetry breaking solution. 
According to the one-matrix model results,  
one may expect that the more physical branch of two-cut solution should interpolate 
between the asymmetric one-cut and the symmetric two-cut solutions, 
such as those in \eqref{1-cut-breaking-kmax-2} and \eqref{2-cut-symmetric-solution-kmax-2} 
for $k_\text{max}=2$. 
However, as the numbers of solutions do not match, 
it is impossible that all the asymmetric one-cut solutions are smoothly connected to the symmetric two-cut solutions. 
In figure \ref{fig:2mm-2-cut-contour}, 
we present the real solutions of the two-cut ansatz in the relevant region. 
\footnote{There are many real solutions that are not in this region.} 
For a given $G_1$,  the free parameters are determined by 
the constraint associated with $G_1$,  
the normalization condition $G_{0}=1$,  
and the self-consistent constraints in \eqref{breaking-constraint-1}, \eqref{breaking-constraint-2}, \eqref{breaking-constraint-3}, \eqref{breaking-constraint-5}.  
In figure \ref{fig:2mm-2-cut-contour}, 
there are two nearly degenerate solutions in the lowest branch for $|G_1|\leq 3.8$ 
and then three nearly degenerate solutions at $(G_1, G_2)\approx (\pm 3.9, 29.4)$. 
\footnote{For $G_1=\pm 3.88$, the three nearly degenerate solutions are $G_2\approx 29.40713, 29.40766, 29.40774$, which are consistent with the Monte Carlo results $G_1=\pm 3.88(2), G_2=29.41(4)$ in \cite{Jha:2021exo}, as well as the upper limit of the positivity bounds in \cite{Kazakov:2021lel}. 
The Monte Carlo results were obtained from the starting matrix $A,B=I$. 
It is not clear if the fact that the corresponding results are close to the collision point at  $G_1\approx 3.9$ 
is a coincidence or related to the collision phenomena. }
As $|G_1|$ increases, the continuation of the symmetric solution and the middle branch meet and become a pair of complex conjugates.  
In figure \ref{fig:2mm-2-cut-contour-complex}, 
we present the continuations of these three solutions in the complex space. 
The annihilated solutions are weakly complex in $3.9\leq|G_1|\leq 5.27$ and 
become real again for $|G_1|\geq 5.28$. 
For $G_1=\pm 5.329$, the lowest branch gives $G_2\approx 28.9407$, 
which is slightly above the asymmetric one-cut solutions in \eqref{1-cut-breaking-kmax-2}. 
We also find considerably more real two-cut solutions around the asymmetric one-cut solutions. 
It is not clear to us if the small imaginary parts are artifacts of our crude approximation scheme. 
To clarify this point, we need to study the cases of $k_\text{max}>2$, 
but they are computationally more expensive and left for the future. 

\section{Discussion}
\label{sec:discussion}
In conclusion, we have demonstrated the effectiveness of the analytic trajectory bootstrap method in solving the large $N$ two-matrix model with $\text{tr}[A,B]^2$ interaction and quartic potentials. 
Our results provide highly accurate solutions (see e.g. table \ref{table-2mm}) and offer new insights into the behavior of eigenvalue distributions and symmetry-breaking solutions.
While not relying on positivity requirements, our method exceeded the accuracy of the relaxation bootstrap 
by fully utilizing the non-linear loop equations. 
The analytic trajectory bootstrap method thus proves to be 
a promising tool for efficiently solving large $N$ matrix models, 
providing highly accurate solutions and paving the way for further investigations into more complex scenarios.

In section \ref{sec:eigenvalue-distributions}, 
we discussed the distributions of eigenvalues for one-length and two-length Green's functions. 
From the matrix integral perspective, 
the angular dependence has been integrated out and we present the results for the one-length and two-length eigenvalue densities. 
Our approach and results may shed light on a better understanding of these angular integrals. 
For $g=0$, the two-matrix model is analytically solvable \cite{Hoppe:1982en, Kazakov:1998ji}, 
even though there are infinitely many base moments, i.e., $\<\text{Tr}\,A^{2j}\>$ with $j=0,1,2,\dots$.  
For generic parameters $g\neq 0, h\neq 0$, the additional base moments for the symmetric solution are
$\<\text{Tr}\,A^{4j-2}B^2\>$ with $j=1,2,\dots$, 
so the number of free parameters is just about $1.5$ times that of the $g=0$ case. 
Given the rapid convergence and low computational cost in our study,  
it is possible that  
the two-matrix model \eqref{2mm-definition} is analytically solvable 
in some novel sense or by some novel methods
for $g=h=1$ or, more generally, $g\neq 0, h\neq 0$.

In section \ref{sec:1mm} and section \ref{sec:2mm}, we mainly focus on the real solutions, 
but the analytic trajectory bootstrap method naturally applies to the complex cases, 
which may not be studied by the positivity-based bootstrap methods. 
For instance, some weakly complex solutions are examined in section \ref{sec:symmetry-breaking}. 
In general, the dimension of the search space will be doubled, 
as a complex number is specified by two real numbers, the real and imaginary parts. 
From the algebraic geometry perspective, there are no essential differences between the real and complex solutions, 
i.e., we are not able to directly extract to the real solutions. 
We first find out all the solutions in the complex parameter space and then select the real solutions 
with vanishing imaginary parts. 
It would be useful to develop a systematic homotopy-continuation approach, 
so we can study the continuous families of solutions from some special limits 
with higher symmetries or/and simpler singularity structure, which are easier to solve accurately. 
In quantum field theory, a similar strategy is to first study the fixed points of renormalization group flow
and then the flow between them.
The fixed points as special limits can have 
higher symmetries, such as scale or conformal invariance, and 
simpler analytic structure. 
It would also be interesting to revisit the conformal bootstrap \cite{Poland:2018epd, Hartman:2022zik, Rychkov:2023wsd} 
by expanding the correlators in terms of more sophisticated building blocks 
that can capture both the global analytic structure and the leading singularities in a systematic manner. 
\footnote{The conformal correlators are analogous to the generating functions of the matrix models. 
See \cite{Li:2023tic} for a recent conformal bootstrap study of the 3D Ising CFT, which shares the same spirit as the present work. }

We plan to apply the analytic trajectory bootstrap to 
more complicated  multi-matrix models, especially the supersymmetric theories.  
Some prominent targets are the Ishibashi-Kawai-Kitazawa-Tsuchiya matrix model \cite{Ishibashi:1996xs}, 
the Banks-Fischler-Shenker-Susskind matrix quantum mechanics \cite{Banks:1996vh},  
and the Berenstein-Maldacena-Nastase matrix quantum mechanics \cite{Berenstein:2002jq}. 
In lattice gauge theory, it is natural to consider the large length limit,  
as the Wilson loop serves as a nonlocal order operator for confinement. 
In a deconfined phase,  the expectation value of a large Wilson loop is expected to obey the perimeter law. 
In a confined phase, the leading behavior of a large Wilson loop should follow the area law 
due to the linearly growing potential $V(r)\sim \s r$ between static charges, 
where  $\s$ is the string tension. 
Here a confining flux tube furnishes a semi-classical picture for the large length limit. 
It would be fascinating to bootstrap the phases of gauge theories 
by the large length expansion and the analytic continuation in the loop length, 
which give rise to various types of trajectories of Wilson loops. 
The diverse choices for the loop configurations should lead to 
even more intricate intersections of the analytic trajectories than 
the case of the two-matrix model in figure \ref{fig:one-length-trajectories}. 
Another natural type of observables is associated with   
a large power of identical plaquettes, i.e., the large multiplicity limit.

\section*{Acknowledgments}
I would like to thank Zechuan Zheng for helpful discussions. 
This work was supported by 
the Natural Science Foundation of China (Grant No. 12205386) and 
the Guangzhou Municipal Science and Technology Project (Grant No. 2023A04J0006).

\appendix

\section{Analytic ansatz from matrix integrals}
\label{appendix:ansatz-integral}
For the one-matrix model, a $\mathbb Z_2$-symmetric 1-cut solution gives
\be
G_n=\<\text{Tr}\,M^n\>=\int_{-z_\ast}^{z_\ast} dz\,z^n \rho(z)\,,
\ee
where the  density of the eigenvalue distribution is given by
\be
\rho(z)=-\frac{1}{2\pi i}\left[R(z+i0)-R(z-i0)\right]
=-\frac{1}{\pi i}\sum_{k=0}^{k_\text{max}}\, a_k\,(z^2-z_\ast^2)^{\frac 1 2+k}\,.
\ee

\label{sec:ansatz-integral}
\begin{figure}[h]
	\centering
		\includegraphics[width=1\linewidth]{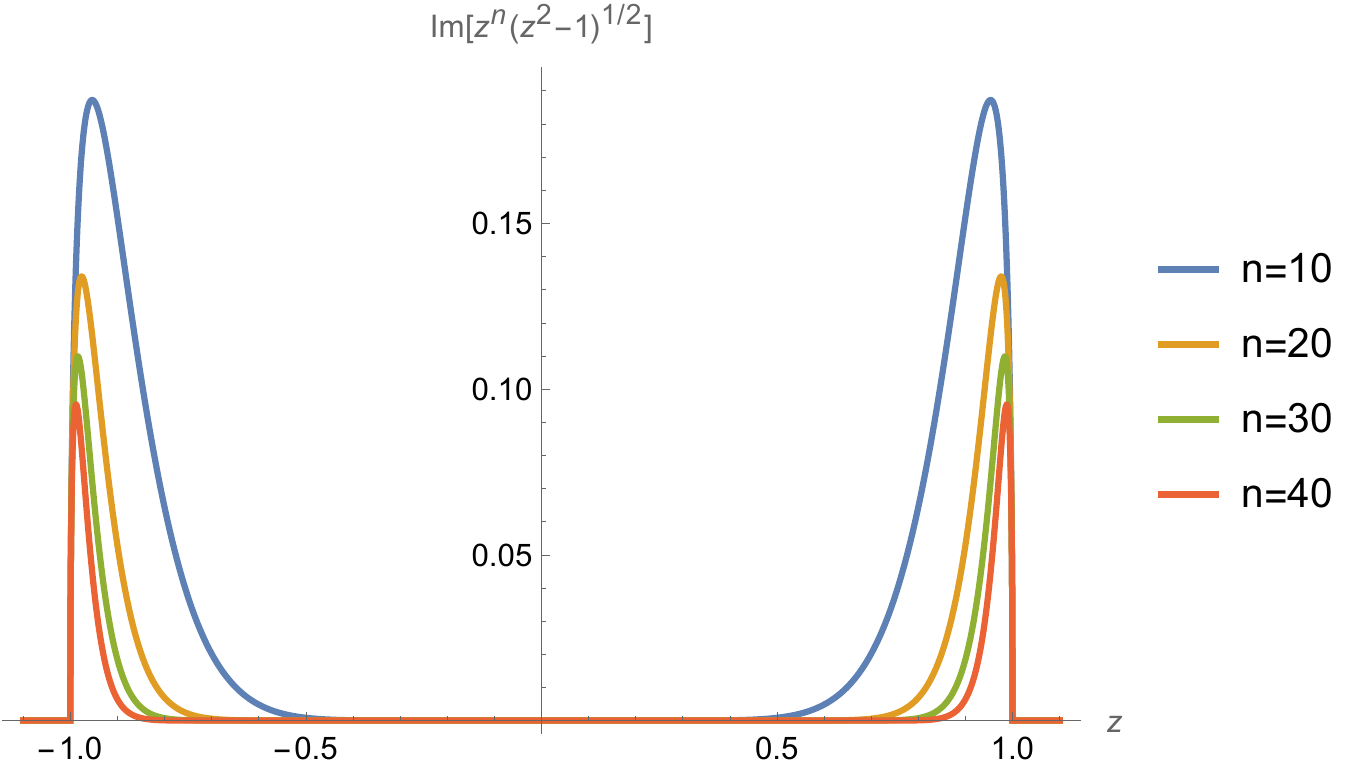}
	\caption{
	As $n$ increases, the dominant contributions in \eqref{z-integral} localize near the boundary of the integration interval. At large $n$, the maxima of $\text{Im}[z^n(z^2-1)^{1/2}]$ exhibit an $n^{-1/2}$ decay, while the widths are proportional to $n^{-1}$. Here we choose $z^2_\ast=1$, so the $z_\ast^{n+2}$ dependence is not visible.  }
	\label{large-n-integral}
\end{figure}

According to the basic integral
\be
\int_{-z_\ast}^{z_\ast} dz\,z^n (z^2-z_\ast^2)^{\frac 1 2+k}
=-\pi i\frac{1+(-1)^n}2
\frac{ (-1/2-k)_{k+1}}{(k+1)!}z_\ast^{2(k+1)}\frac{(1/2)_{n/2}}{(k+2)_{n/2}}\,z_\ast^n\,,
\label{z-integral}
\ee
we obtain the analytic ansatz \eqref{1mm-Gn-ansatz} 
\be
G_n=\frac{1+(-1)^n}2\sum_{k=0}^{k_\text{max}}
\left(a_k\frac{ (-1/2-k)_{k+1}}{(k+1)!}z_\ast^{2(k+1)}\right)\frac{(1/2)_{n/2}}{(k+2)_{n/2}}\,z_\ast^n\,.
\ee
At large $n$, the leading asymptotic behavior of $G_n$ is
\be
G_n\sim  \frac{1+(-1)^n}2  \frac{(-2)a_0}{\pi^{1/2}}\,n^{-3/2}z_\ast^{n+2}\quad
(n\rightarrow \infty),
\label{Gn-large-n}
\ee
which is associated with the $k=0$ term. 

The leading $n$ dependence can also be deduced from the integral perspective. 
At large $n$, the leading contributions of \eqref{z-integral} are 
localized near the largest eigenvalues $|z|\approx z_\ast$,  
which is shown in figure \ref{large-n-integral}. 
They are not precisely at $z^2=z_\ast^2$ because the eigenvalue density vanishes at the largest eigenvalue. 
We can determine the approximate location by $\frac d {dz}(z^n(z^2-z_\ast^2)^{1/2})=0$, 
which gives the positions $z=\pm z_\ast (1+1/n)^{-1/2}\sim (1-1/(2n))z_\ast$ 
and the maximum integrand $\text{Max}[z^n(z^2-z_\ast^2)^{1/2}]\sim i(e \,n)^{-1/2}z_\ast^{n+1}$. 
We can also estimate the width of the localized region by $\frac {d^2} {dz^2}(z^n(z^2-z_\ast^2)^{1/2})=0$,
which gives $z\sim \pm (1-(1+\sqrt{2})/(2n))z_\ast$,  
so the width is of order $n^{-1}z_\ast$ and consistent with the location of the maximum integrand. 
The leading $n$ dependence of $G_n$ can be deduced from a product of the height and width
$(n^{-1/2}z_\ast^{n+1})(n^{-1}z_\ast )=n^{-3/2}z_\ast^{n+2}$, 
which is consistent with the asymptotic behavior in \eqref{Gn-large-n}.  

For multi-length Green's functions, 
it is natural to expect that the leading contributions in the large length limit also localize 
around the largest eigenvalues $|z|=z_\ast$, 
which correspond to the edges of the support of the eigenvalue distributions.  
Therefore, the large length limit of the multi-length Green's functions is dominated by a factorized term,  
such as the two-length case in \eqref{Gnn-kmax-0}. 
\footnote{In situations with more symmetry, 
the observables should be compatible with stronger symmetry constraints. 
For SO(D) symmetry, it is natural to choose independent singlets as the building blocks 
if non-singlets all vanish.  }

\end{document}